\documentclass[referee]{chjaa}          

\usepackage{graphicx,times}             
\input{epsf.sty}                        
\input{psfig.sty}                       

\begin{document}

   \title{Third Order Effect of Rotation on Stellar Oscillations of a B Star
}

   \volnopage{Vol.0 (200x) No.0, 000--000}      
   \setcounter{page}{1}           

   \author{K. Karami
      \inst{1,2,3}\mailto{}
            }
   \institute{Department of Physics, University of Kurdistan, Pasdaran St.,
             P. O. Box 66177-15175, Sanandaj, Iran\\
             \email{Karami@iasbs.ac.ir}
        \and
             Research Institute for Astronomy $\&$ Astrophysics of Maragha
(RIAAM), P. O. Box 55134-441, Maragha, Iran\\
        \and
             Institute for Advanced Studies in Basic Sciences (IASBS), Gava Zang, P. O. Box 45195-1159, Zanjan, Iran
\\
          }

   \date{Received~~2001 month day; accepted~~2001~~month day}

   \abstract{
The present paper aims at investigating the effect of rotation up to
third order in the angular velocity of a star on p and g modes,
based on the formalism developed by Soufi et al. (1998). Our
ultimate goal is the study of oscillations of $\beta$ Cephei stars
which are often rapidly rotating stars. Our study shows that the
third-order perturbation formalism presented by Soufi et al. (1998)
must be corrected for some missing terms and some misprints in the
equations. As a first step in our study of $\beta$ Cephei stars, we
quantify by numerical calculations the effect of rotation on the
oscillation frequencies of a uniformly rotating zero-age
main-sequence star with 12~$M_\odot$. For an equatorial velocity of
100~$\rm km \, s^{-1}$, it is found that the second- and third-order
corrections for $(l,m)=(2,2)$, for instance, are of order of
0.01$\%$ of the frequency for radial order $n=6$ and reaches up to
0.5$\%$ for $n=14$.
   }

   \authorrunning{K. Karami}            
   \titlerunning{Third order effect of rotation}  

   \maketitle

%
%
\section{Introduction}           
\label{sect:intro}

Pulsating stars on the upper main sequence, and particularly
$\delta$ Scuti and $\beta$ Cephei stars, are rapid rotators as well
as being multimode pulsators. The ratio $\epsilon=\Omega/\omega$ of
the rotation rate $\Omega$ to the typical frequency of oscillations
$\omega$ seen in these stars is no longer a small quantity as it is
for e.g. the Sun. These stars have typically equatorial velocities
$\sim 100\ {\rm km \, s^{-1}}$, and oscillation periods from half to
a few hours which implies $\epsilon\sim 0.1$, whereas for the Sun
$\epsilon\sim 10^{-4}$. The effect of rotation on stellar structure
and stellar oscillations is usually calculated through a
perturbation analysis in which $\epsilon$ is the small parameter.
Whereas for the Sun a first-order perturbation analysis is
sufficient, given the accuracy of observed oscillation frequencies,
for these stars it is not. In order to achieve the full potential of
asteroseismology for testing of models for upper main sequence stars
a more careful treatment of the effect of rotation on oscillation
frequencies is required.

Rotation not only modifies the structure of the star but also
changes the frequencies of normal modes. It removes mode degeneracy
creating multiplets of modes. If the rotational angular velocity,
$\Omega$, does not have any latitudinal dependence, and the rotation
is sufficiently slow, the multiplets show a Zeeman-like equidistant
structure. At faster rotation rates non-negligible quadratic effects
in $\Omega$ cause the position of the centroid frequency of
multiplets to shift with respect to that of a non-rotating model of
the same star. See Karami et al. (2003).

Our long term goal is to study the oscillation properties of rapidly
rotating $\beta$ Cephei star. In this paper, we start with the study
of rotation up to third order in the angular velocity of the star on
p and g normal modes. To do this we use the third-order perturbation
formalism according to Soufi et al. (1998), hereafter S98, but
correct some misprints and missing terms in some of their equations.
We carry out numerical calculations for the frequency corrections
for a stellar zero-age main-sequence (ZAMS) model with mass $M = 12
\, M_{\odot}$, $M_\odot$ being the solar mass. Section 2 gives a
brief overview of previous relevant work on rotational perturbation
theory for stellar oscillations. Section 3 shows the equations for
the effect of rotation on oscillations up to third order and
discusses the zeroth-order eigensystem. Section 4 presents
corrections to eigenfrequencies and coupling coefficients. The
numerical results are presented in section 5. Section 6 is devoted
to concluding remarks. A summary of the differences between the
present formulation and that of S98 is given in Appendix
\ref{Appendix-summary}. Lengthy formulae are collected in Appendix
\ref{Appendix-Dkq}.
The hermiticity properties of an oscillating rotating system are
discussed in Appendix~\ref{Appendix-Hermiticity}.

\section{Previous rotational perturbation analysis}
Simon (1969) and Saio (1981) studied the frequency corrections due
to rotation up to second order for polytropes. Chlebowski (1978)
calculated the same corrections for white dwarf models. Gough \&
Thompson (1990) studied the effects of rotation and magnetic field
on stellar oscillations up to second order. They investigated the
linearized adiabatic oscillation equation in a rotating frame under
the Cowling approximation and without viscous and resistive
dissipative forces. They considered only an axisymmetric magnetic
field, but allowed for its axis not to coincide with the rotation
axis. They concluded that rotation and a magnetic field not only
split the degenerate frequency multiplets but also shift the central
frequency of each multiplet. The shift arises both from the direct
effect of the perturbed inertial and Lorentz forces on the waves,
and also because the unperturbed centrifugal and Lorentz forces
change the structure of the star, which is no longer spherically
symmetric. Gough \& Thompson (1990) succeeded in formulating the
differential equations governing the oscillatory motions in a form
that is hermitian. The hermiticity of the oscillation equations for
the rotating star was preserved by means of an appropriate mapping
of each point in the distorted model to a corresponding point in the
spherically symmetric stellar model. Although their main attention
was on the effect of a magnetic field on the normal modes, they also
performed some numerical calculations of the effect of rotation for
three different, latitudinally independent, angular velocity
profiles $\Omega(r)$. They found that the effect of the second-order
centrifugal distortion changes little with spherical harmonic
degree, $l$, and can be approximated well by the asymptotic
estimate. Also, for low $l$ the second-order correction due to the
advection term is negligible compared with the second-order
centrifugal distortion and only for $l \geq 50$ are the two
comparable in magnitude.

Dziembowski \& Goode (1992) derived a formalism for calculating the
effect of differential rotation on normal modes of rotating stars up
to second order. They considered angular-velocity profiles with both
radial and latitudinal dependency and found that at faster rotation
non-negligible quadratic effects in $\Omega$ cause a departure from
equidistant splitting. They also obtained generalized asymptotic
formulae for g-mode splitting for which the Coriolis term is
included in the zero-order treatment. These asymptotic results are
relevant for white dwarfs and $\delta$ Scuti stars. Dziembowski \&
Goode (1992) concluded that for solar oscillations the second-order
effects are dominated by distortion for $l<500$.

Soufi et al. (1998) extended the formalism of Dziembowski \& Goode
(1992) up to third order for a rotation profile that is a function
of radius only. Their analysis has two advantages compared with
previous investigations. By taking into account parts of the effects
of the Coriolis force in the zero-order system, the eigenvalue
problem for stellar oscillations can be solved up to cubic order
without having to solve successive equations for the eigenfunctions
at each order. Also the usual $m$-degeneracy occurring in the
absence of rotation is removed at the lowest order. S98 found that
near-degenerate coupling due to rotation only occurs between modes
with either the same degree $l$ (and different radial orders) or
with modes which differ in degree by 2. The first case involves
modes in avoided crossings. The second case concerns modes that have
close enough frequencies in the non-rotating model to be shifted
into resonance if the rotation is sufficiently rapid. In general
they showed that the total coupling comes from three distinct
contributions: the Coriolis contribution, the
non-spherically-symmetric distortion, and a coupling term which
involves a combination of these two effects.

Sobouti (1980) studied the normal modes of rotating fluids up to
$O(\Omega^2)$. He argued that the p modes allow a perturbation
expansion in $\Omega$, whereas this is not the case for the g modes.
For, from a mathematical point of view, the condition for a
perturbation series to converge is that the perturbing operator
remain smaller than the unperturbed operator throughout the Hilbert
space spanned by the normal modes; this condition is not met by g
modes.

Sobouti \& Rezania (2001) considered the toroidal modes of rotating
fluids, and showed that a) At $O(\Omega)$ the neutral toroidal
motions of the non-rotating fluid organize themselves into a
sequence of modes with a definite $(l,m)$ symmetry. Their radial
degeneracy, however, persists at this order. b) Coupling of a given
toroidal mode of $(l,m)$ symmetry with $(l\pm2,m)$ toroidal and with
$(l\pm1,m)$ poloidal modes, as well as the removal of the radial
degeneracy come about at $O(\Omega^2)$.

Result of calculation of frequency corrections up to third order
were presented for models of $\delta$-Scuti stars by Goupil et al.
(2001), Goupil \& Talon (2002), Pamyatnykh (2003), and Goupil et al.
(2004). Daszy\'{n}ska-Daszkiewicz et al. (2002) studied the effects
of mode coupling due to rotation on photometric parameters
(amplitude and phase) of stellar pulsations. They reconfirmed the
conclusion of S98 that the most important effect of rotation is
coupling between close frequency modes of spherical harmonic degree,
$l$, differing by 2 and of the same azimuthal order, $m$. They
presented some numerical results for a sequence of $\beta$ Cephei
star models with uniform rotation and for two- and three-mode
couplings. Their calculations were carried out to cubic order in the
ratio of rotation to pulsation frequency, according to the
third-order formalism of S98. They concluded that due to the
increasing effect of centrifugal distortion with the mode frequency,
the coupling between acoustic modes is stronger than between gravity
modes.

Reese et al. (2006) studied the effects of rotation due to both the
Coriolis and centrifugal accelerations on pulsations of rapidly
rotating stars by a non-perturbative method. They showed that the
main differences between complete and perturbative calculations come
essentially from the centrifugal distortion. Su\'{a}rez et al.
(2006) obtained the oscillation frequencies include corrections for
rotation up to second order in the rotation rate for $\delta$ Scuti
star models.

\section{Third order perturbation formalism}
Following Unno et al. (1989), the equilibrium state of a rotating
star can be characterized by a velocity field
\begin{equation}
\mathbf{v}_0=\mathbf{\Omega}\times\mathbf{r}=\Omega r
\sin\theta\mathbf{e}_\varphi \; , \label{velfld}
\end{equation}
where $\Omega$ denotes the angular velocity. The rotation axis of
the star lies in the $\theta=0$ axis of a spherical coordinate
system $(r,\theta,\varphi)$. $\Omega$ is assumed to be independent
of latitude and can be written as:
\begin{equation}
\mathbf{\Omega}=\Omega(r)\mathbf{e}_z=\bar{\Omega}[1+\eta(r)]\mathbf{e}_z
\; , \label{Omeass}
\end{equation}
where $\bar{\Omega}$ is the mean rotation rate. The stationary
equation of motion in an inertial frame of reference is:
\begin{equation}
-(\mathbf{v}_0\cdot\nabla)\mathbf{v}_0=\frac{\nabla
p}{\rho}+\nabla\phi \; , \label{hydeq}
\end{equation}
where $p$, $\rho$, and $\phi$ are the pressure, density and
gravitational potential, respectively. One can show that the left
hand side of Eq.~(\ref{hydeq}) is equal to the centrifugal
acceleration, $\mathbf{F}$, in a corotating frame as follows:
\begin{equation}
\mathbf{F}=-\mathbf{\Omega}\times(\mathbf{\Omega\times\mathbf{r}})
=r\Omega^2\sin \theta \mathbf{e}_s
=-(\mathbf{v}_0\cdot\nabla)\mathbf{v}_0 \; ,
\end{equation}
where $\mathbf{e}_{z}$, and $\mathbf{e}_{s}=\sin\theta
\mathbf{e}_{r}+\cos \theta \mathbf{e}_{\theta}$ are unit vectors in
cylindrical coordinates ($s,\varphi,z$). See Tassoul (2000).
\subsection{Equilibrium structure of rotating stars}
The stationary equation of motion, Eq.~(\ref{hydeq}), is solved
following Chandrasekhar (1933) and Chandrasekhar \& Lebovitz (1962)
by expanding the equilibrium quantities in terms of Legendre
polynomials as
\begin{equation}
f(r,\theta)=\tilde{f}(r)+\epsilon^{2}f_2=\tilde{f}(r)+
\epsilon^{2}f_{22}(r)P_2(\cos\theta) \; , \label{LegenExpan}
\end{equation}
where $P_2(\cos\theta)=3/2\cos^2\theta-1/2$ is the second Legendre
polynomial and $f$ can be $p$, $\rho$ or $\phi$. For a rotation rate
that is a function of $r$ only, higher-order multipole moments
(i.e., $P_4, P_6,..$) need not be considered.

\subsubsection{Spherically symmetric distortion} The spherically
symmetric part of the equilibrium structure can be obtained by
substituting Eq.~(\ref{LegenExpan}) in the $\theta$-independent part
of the radial component of Eq.~(\ref{hydeq}). The result is
\begin{equation}
\frac{d\tilde{p}}{dr}=-\tilde{\rho}g_{\rm e} \; , \label{sphsymp}
\end{equation}
where the effective gravity is $g_{\rm e}$:
\begin{equation}
g_{\rm e}=\tilde{g}-\frac{2}{3}r\Omega^2 \; , \label{effgrav}
\end{equation}
with
\begin{equation}
\tilde{g}=\frac{d\tilde{\phi}}{dr}=\frac{GM_r}{r^2} \; ,
\label{sphsymgrav}
\end{equation}
and $M_r$ is the mass confined within radius $r$.
Eq.~(\ref{sphsymp}) is similar to the equation of hydrostatic
equilibrium of a non-rotating star but modified by including the
spherically symmetric part of the centrifugal force. Note that the
other equations governing the quantities of internal structure do
not change; hence as in S98, following the approach described by
Kippenhahn \& Weigert (1994) a standard evolutionary code only
modified according to Eq.~(\ref{sphsymp}) can be used to follow the
evolution. Of course, in order to compute a model at a given age,
the profile of the rotation rate is needed which requires the
knowledge of its temporal evolution.

\subsection{Non-spherically symmetric distortion}
Substituting Eq.~(\ref{LegenExpan}) in the $\theta$-dependent parts
of the radial and the tangential components of Eq.~(\ref{hydeq}) and
neglecting $O(\epsilon^4)$ terms yields the
non-spherically-symmetric part of the equilibrium structure as:
\begin{equation}
p_{22}=-\tilde{\rho}r^2(\bar{\Omega}/\epsilon)^2u_2 \; ,
\label{p22term}
\end{equation}
\begin{equation}
\rho_{22}=\frac{\tilde{\rho}r(\bar{\Omega}/\epsilon)^2}{\tilde{g}}
\left(\frac{d\ln\tilde{\rho}}{d\ln r}u_2+b_2 \right) \; ,
\label{rho22term}
\end{equation}
where
\begin{equation}
\begin{array}{rl}
u_2 &=\frac{\displaystyle \phi_{22}}{\displaystyle r^2}
(\bar{\Omega}/\epsilon)^{-2} +
\frac{1}{3}(1+\eta_2) \; , \\
b_2 &=\frac{1}{3}r\frac{\displaystyle d\eta_2}{\displaystyle dr} \; , \\
\end{array}
\label{ubdefs}
\end{equation}
with $\eta_2=\eta(\eta+2)$. The perturbed gravitational potential
$\phi_{22}$ satisfies the perturbed Poisson equation:
\begin{equation}
\frac{1}{r^2}\frac{d}{dr}\left(r^2\frac{d\phi_{22}}{dr}\right)-
\frac{6}{r^2}\phi_{22}=4\pi G\rho_{22} \; . \label{phi22term}
\end{equation}
Equation~(\ref{phi22term}) can be solved by numerical integration
and with the appropriate boundary conditions: $\phi_{22}\propto r^2$
at the centre of the star, and $\phi_{22}\propto r^{-3}$ at the
surface of the star. The relations (\ref{p22term}) to
(\ref{phi22term}) are identical to Eqs. (77)--(80) (with $k$=1) of
Dziembowski \& Goode (1992) and also with Eqs. (15)--(17) of S98.

\subsection{Zeroth-order eigensystem}
\label{Section-zeroth}
Following S98, we include parts of the Coriolis and non-spherical
distortion effects in the zero-order eigensystem. This yields
eigenfrequencies $\omega_0$ of eigenmodes which are no longer
$m$-degenerate, even at zero order. This way of building the
zero-order eigensystem and the associated basis of eigenmodes
enables one to solve the eigenvalue problem up to cubic order
without having to solve the successive equations for the
eigenfunctions at each order.

A zero-order mode $\xi_0$ is defined by
\begin{equation}
\xi_0=\xi_{p0}+\epsilon\xi_{\rm t1} \; , \label{zerordmod}
\end{equation}
where the poloidal $\xi_{p0}$ and toroidal $\xi_{\rm t1}$
eigenfunctions are characterized by a single spherical harmonic, as
in the case of a non-rotating model, and are given by:
\begin{equation}
\begin{array}{rl}
\xi_{p0} &=r(yY_l^{m}+z\nabla_{\rm H} Y_l^{m}) \\
\xi_{\rm t1}
&=r\frac{\displaystyle\bar{\Omega}}{\displaystyle\hat{\omega}_{0}}
(\tau_{l+1}\mathbf{e}_r
\times\nabla_{H}Y_{l+1}^{m}+\hat{\tau}_{l-1}\mathbf{e}_r\times
\nabla_{H}Y_{l-1}^{m}) \; , \\
\end{array}
\label{zerordpoltorxi}
\end{equation}
where $\nabla_{\rm H}
\equiv(\frac{\partial}{\partial\theta}\mathbf{e}_\theta+
\frac{1}{\sin\theta}\frac{\partial}{\partial\varphi}\mathbf{e}_\varphi)$
is the horizontal part of the gradient operator and
$\hat{\omega}_0=\omega_0+m\Omega$.

According to Eq. (39) in S98 the eigenfrequency $\omega_0$ can be
rewritten as:
\begin{equation}
\omega_0=\omega_0^{(0)}+\omega_1 \; , \label{ome00exp}
\end{equation}
where $\omega_0^{(0)}$ is the usual normal-mode frequency of the
system when rotation is absent, and
\begin{equation}
\omega_1=m\bar{\Omega}(C_{\rm L}-1-J_1) \; . \label{ome1def}
\end{equation}
The quantities $C_{\rm L}$ and $J_1$ are given by Eq. (40) in S98
and references therein. Note that, following S98, we assume a
dependence of the perturbations on time $t$ and longitude $\varphi$
as $\propto \exp[i (m \varphi + \omega t)]$, such that prograde
modes have $m < 0$.

The frequency $\omega_0$ includes the first-order correction due to
rotation, $\omega_1$, as well as the second-order correction due to
spherically symmetric distortion. Also note that $\omega_1$ is the
usual first-order frequency shift due to the Coriolis force and can
be rewritten as:
\begin{equation}
\omega_1=-\frac{\displaystyle m}{\displaystyle I}
\int\Omega(r)\left[ y^2+(\Lambda-1)z^2-2yz\right]r^4\tilde{\rho}dr
\; , \label{ome1calc}
\end{equation}
where $I$ is the mode inertia,
\begin{equation}
I=\langle\xi_{p0}\mid\tilde{\rho}\xi_{p0}\rangle=\int
dr\tilde{\rho}r^4(y^2+\Lambda z^2) \; , \label{modeiner}
\end{equation}
with $\Lambda=l(l+1)$.

\subsection{Poloidal eigenfunctions}
Following Unno et al. (1989), we define the dimensionless variables
$y_t$, $v$ and $w$ as
\begin{equation}
y_t\equiv\frac{1}{g_{\rm e} r}\left(\tilde{\phi}' +
{\tilde{p}'\over\tilde{\rho}} \right) \; ,~~~~
v\equiv\frac{\tilde{\phi}'}{g_{\rm e} r} \;
,~~~~~w\equiv\frac{1}{g_{\rm e}} \frac{d\tilde{\phi}'}{dr} \; .
\label{dimlsyt}
\end{equation}
Then as in S98, the expression for the poloidal components is
obtained as
\begin{equation}
\begin{array}{rl}
r\frac{\displaystyle dy}{\displaystyle dr}&=
(V_g-3+h_1)y+(\zeta-V_g)y_t+V_gv \\
&=\lambda-3y+\Lambda z \; , \\
\end{array}
\label{ydrdifeq}
\end{equation}
\begin{equation}
r\frac{dy_t}{dr}=
\left(C_{r}\hat{\sigma}^{2}-A-\frac{h_{1}^{2}}{\zeta} \right)y+
(A+1-U-\chi-h_{1})y_t-Av \; , \label{ytdrdifeq}
\end{equation}
\begin{equation}
r\frac{\displaystyle dv}{\displaystyle dr}=(1-U-\chi)v+w \; ,
\end{equation}
\begin{equation}
r\frac{\displaystyle dw}{\displaystyle dr}= \frac{\displaystyle
UA}{\displaystyle 1-\sigma_r}y+ \frac{\displaystyle
UV_g}{\displaystyle 1-\sigma_r}y_t+
\left(\Lambda-\frac{\displaystyle UV_g}{\displaystyle 1-\sigma_r}
\right)v-(U+\chi)w \; , \label{wdrdifeq}
\end{equation}
\begin{equation}
\Lambda z-\zeta y_t-h_1y=0 \; . \label{ampleqn1}
\end{equation}
Here $\sigma^2={R^3}\omega^2/{GM}$ is the square of the
dimensionless oscillation frequency and
\begin{equation}
\hat{\sigma}\equiv\sigma+m\sigma_\Omega \; ,~~~~
\sigma_\Omega\equiv\frac{\Omega}{\sqrt{GM/R^3}} \; ,
\end{equation}
\begin{equation}
A\equiv\frac{1}{\Gamma_1}\frac{d\ln \tilde{p}}{d\ln r}-\frac{d\ln
\tilde{\rho}}{d\ln r} \; ,~~~~ V_g\equiv
-\frac{1}{\Gamma_1}\frac{d\ln\tilde{p}}{d\ln r} \; , \label{AUdefs}
\end{equation}
\begin{equation}
U\equiv\frac{d\ln M_r}{d\ln r} \; ,~~~~
\sigma_r\equiv\frac{2}{3}r\Omega^2/\tilde{g} \; , \label{AUdefs}
\end{equation}
\begin{equation}
C\equiv\frac{(r/R)^3}{(M_r/M)} \; ,~~~~
C_r\equiv\frac{C}{1-\sigma_r} \; , \label{CrCsigrdef}
\end{equation}
\begin{equation}
\lambda\equiv V_g(y-y_t+v) \; ,~~~~ \alpha\equiv
2m\frac{\sigma_\Omega}{\hat{\sigma}} \; ,
\end{equation}
\begin{equation}
\chi\equiv\frac{2r\Omega^2}{3 g_{\rm e}} \left(U-3-\frac{d\ln
\Omega^2}{d\ln r} \right) \; , \label{chidef}
\end{equation}
\begin{equation}
\zeta\equiv\frac{\displaystyle\Lambda}{\displaystyle\Lambda-\alpha}
\frac{\displaystyle\Lambda}{\displaystyle C_r\hat{\sigma}^2} \;
,~~~~
h_1\equiv\frac{\displaystyle\Lambda\alpha}{\displaystyle\Lambda-\alpha}
\; . \label{zetah1def}
\end{equation}
Here $M$ and $R$ are the stellar mass and radius respectively and
$G$ is the gravitational constant.

For radial oscillations ($l=\Lambda=0$) the Poisson equation for
$\xi_{p0}$, Eq. (\ref{phi22term}), reduces to
\begin{equation}
\frac{d\tilde{\phi}'}{dr}=-\tilde{g}Uy \; . \label{poiseqrad}
\end{equation}
Thus $\tilde\phi'$ can be eliminated from the problem; instead of
Eq.~(\ref{dimlsyt}) we define the dimensionless variable $y_t$ in
this case as
\begin{equation}
y_t=\frac{\tilde{p}'}{g_{\rm e} r\tilde{\rho}} \; ,
\label{dimlsytrad}
\end{equation}
obtaining
\begin{equation}
r\frac{dy}{dr}=(V_g-3)y-V_gy_t \; , \label{ydrdifeqrad}
\end{equation}
\begin{equation}
r\frac{dy_t}{dr}= \left(C_r\hat{\sigma}^{2}-A+\frac{U}{1-\sigma_r}
\right)y+ (A+1-U-\chi)y_t \; . \label{ytdrdifeqrad}
\end{equation}
For radial oscillations there is no transverse component, i.e.,
$z=0$.

\subsection{Toroidal eigenfunctions}
Following again S98 the components $\tau$, $\hat{\tau}$ of the
toroidal part, $\xi_{\rm t1}$, for a mode $k\equiv (n_k,l_k,m_k)$,
are obtained as follows:
\begin{equation}
\begin{array}{rl}
\tau_{k+1} & \equiv \tau_{n_k,l_k+1,m_k} \\
&=i\frac{\displaystyle\beta_{k+1}}{\displaystyle(\Lambda_{l_k+1}-\alpha_{k})}
\left(2P_k+3m_k \frac{\displaystyle \bar{\Omega}}{\displaystyle
\hat{\omega}_0}d_k
\right) \; ,\\
\end{array}
\end{equation}
\begin{equation}
\begin{array}{rl}
\hat{\tau}_{k-1}&=\hat{\tau}_{n_k,l_k-1,m_k} \\
&=i\frac{\displaystyle\beta_{k}}{\displaystyle(\Lambda_{l_k-1}-\alpha_{k})}
\left(2\hat{P}_k+ 3m_k\frac{\displaystyle
\bar{\Omega}}{\displaystyle \hat{\omega}_0}d_k\right)
\; , \\
\end{array}
\end{equation}
with
\begin{equation}
P_{k}=(1+\eta)(l_k+2)(-y_k+l_k z_k) \; ,
\end{equation}
\begin{equation}
\hat{P}_{k}=(1+\eta)[l_k-1)(y_k+(l_k+1)z_k] \; ,
\end{equation}
\begin{equation}
\beta_k=\sqrt{\frac{(l_k^2-m_k^2)}{4l_k^2-1}} \; ,
\label{betafacdef}
\end{equation}
$\beta_{k+1}$ being defined similarly, but with $l_k$ replaced by
$l_k + 1$, and
\begin{equation}
d_k=\left(\frac{g_{\rm e}}{\tilde{g}}v_k+y_k-C\hat{\sigma}_k^2z_k
\right) \left(\frac{d\ln \tilde{\rho}}{d\ln
r}u_2+b_2\right)+\lambda_ku_2 \; . \label{dfacdef}
\end{equation}

\section{Correction to eigenfrequencies and coupling
coefficients}\label{Hkq}

As discussed in S98, in general the correction to the
eigenfrequencies must take into account coupling between nearby
modes. Thus we need to evaluate coupling coefficients between
different modes. However, in the case of an isolated mode, $k$, the
correction $\omega_{\rm c}$, to the eigenfrequency
$\omega=\omega_0+\omega_{\rm c}$, is obtained from
\begin{eqnarray}
\omega_{\rm c}&=&
\frac{H_{kk}}{2I\omega_0^{(0)}}\equiv\mathcal{H}_{kk} \nonumber \\
&=&\omega^{\rm T}+\omega^{\rm D}+\omega^{\rm C}+O(\epsilon^4) \; ,
\label{omecorcal}
\end{eqnarray}
where $\mathcal{H}_{kq}$ is the matrix of the coupling coefficients
and is defined as
\begin{equation}
\mathcal{H}_{kq}\equiv\frac{H_{kq}}{2J_{kq}}=
\mathcal{T}_{kq}+\mathcal{D}_{kq}+\mathcal{C}_{kq}+O(\epsilon^4) \;
, \label{calHmatdef}
\end{equation}
and $J_{kq}=\sqrt{\omega_{0k}^{(0)}\omega_{0q}^{(0)}I_kI_q}$. Note
that for an isolated mode each contribution $\omega^{\rm
T}=\mathcal{T}_{kk}$, $\omega^{\rm D}=\mathcal{D}_{kk}$ and
$\omega^{\rm C}=\mathcal{C}_{kk}$ to the frequency correction arises
from the corresponding diagonal term of the interaction matrix
$\mathcal{H}_{kk}$.

In order to obtain expressions for the different contributions to
the total coupling coefficients $\mathcal{H}_{kq}$, see
Eqs.~(B1)--(B2) in S98, it is convenient to define a dimensionless
radius $x$ and density $\bar{\tilde{\rho}}$:
\begin{equation}
\begin{array}{rl}
x &\equiv r/R \; ,\\
\bar{\tilde{\rho}} &\equiv\tilde{\rho}/\rho_{\rm c} \; , \\
\end{array}
\end{equation}
where $\rho_{\rm c}$ is the value of the density at the centre of
the (deformed) star in equilibrium. The second-order perturbation
terms of pressure, density and gravitational potential are also made
dimensionless by defining the following:
\begin{equation}
\begin{array}{rl}
\bar{p}_{22} &\equiv p_{22}/p_{\rm c} \; , \\
\bar{\rho}_{22} &\equiv\rho_{22}/\rho_{\rm c} \; , \\
\bar{\phi}_{22} &\equiv\phi_{22}/\phi_{\rm c} \; , \\
\end{array}
\end{equation}
where $p_{\rm c}=\rho_{\rm c}\phi_{\rm c}$ and $\phi_{\rm
c}=R^2\bar{\Omega}^2$. Oscillation frequencies are made
dimensionless by making use of the dynamical time scale of the star:
\begin{equation}
\begin{array}{rl}
\sigma_{0}^{(0)} &\equiv
\frac{\displaystyle \omega_{0}^{(0)}}{\displaystyle \sqrt{GM/R^3}} \; , \\
\sigma_{\bar{\Omega}} &\equiv
\frac{\displaystyle \bar{\Omega}}{\displaystyle \sqrt{GM/R^3}} \; . \\
\end{array}
\end{equation}
Dimensionless mode inertia terms are:
\begin{equation}
\begin{array}{rl}
\bar{I}_k &\equiv\frac{\displaystyle I_k}{\displaystyle \rho_{\rm c}R^5} \; ,\\
\bar{J}_{kq} &\equiv\frac{\displaystyle J_{kq}}{\displaystyle
\rho_{\rm c}R^5 \sqrt{GM/R^3}} \\
\noalign{\vskip 5pt}
&=\sqrt{\sigma_{0k}^{(0)}\sigma_{0q}^{(0)}\bar{I}_k\bar{I}_q} \; .\\
\end{array}
\end{equation}

\subsection{Coriolis Contribution: $\mathcal{T}_{kq}$}
As in S98 the elements $\mathcal{T}_{kq}$ are obtained as
\begin{equation}
\begin{array}{rl}
\overline{\mathcal{T}}_{kq} & \equiv
\frac{\displaystyle \mathcal{T}_{kq}}{\displaystyle \sqrt{GM/R^3}} \\
& = \overline{\mathcal{T}}_{qk}= \delta_{l_k l_q}\overline{T}^{(1)}+
\delta_{l_k l_q+2}\overline{T}_{kq}^{(2)}+
\delta_{l_k l_q-2}\overline{T}_{qk}^{(2)\ast} \; , \\
\end{array}
\label{calTsepterms}
\end{equation}
where $\ast$ denotes complex conjugate; here the diagonal and
off-diagonal terms are:
\begin{equation}
\begin{array}{rl}
\overline{T}^{(1)}=&\frac{\displaystyle\sigma_{\bar{\Omega}}^2}{\displaystyle
2\bar{J}_{kq}}\biggl\{\displaystyle\int_0^1dx
\bar{\tilde{\rho}}x^4 \times\biggr.\\
&~~~~~~~~\times(\Lambda_{k+1}\tau_{k+1}^{\ast}\tau_{q+1}+
\Lambda_{k-1}\hat{\tau}_{k-1}^{\ast}\hat{\tau}_{q-1})\\
&-\frac{\displaystyle 4m\sigma_{\bar{\Omega}}}{\displaystyle
\sigma_{0k}+\sigma_{0q}}\displaystyle\int_0^1dx
\bar{\tilde{\rho}}x^4(1+\eta)\times\\
&~~~~~~~~\biggl.\times(\tau_{k+1}^{\ast}\tau_{q+1}+
\hat{\tau}_{k-1}^{\ast}\hat{\tau}_{q-1})\biggr\} \; , \\
\end{array}
\end{equation}
\begin{equation}
\begin{array}{rl}
\overline{T}_{kq}^{(2)}=&\frac{\displaystyle\sigma_{\bar{\Omega}}^2}{\displaystyle
2\bar{J}_{kq}}\left\{\displaystyle\int_0^1dx
\bar{\tilde{\rho}}x^4\Lambda_{k-1}\hat{\tau}_{k-1}^{\ast}\tau_{q+1}\right.\\
&\left. -\frac{\displaystyle 4m\sigma_{\bar{\Omega}}}{\displaystyle
\sigma_{0k}+\sigma_{0q}}\displaystyle\int_0^1dx
\bar{\tilde{\rho}}x^4(1+\eta)\hat{\tau}_{k-1}^{\ast}\tau_{q+1}\right\} \; . \\
\end{array}
\end{equation}
We recall that some effects of distortion are already included in
the toroidal zeroth-order system. Therefore distortion contributes
indirectly through the toroidal part of the eigenfunctions, i.e.,
through the components $\tau$ and $\hat{\tau}$.

Note that the case $k\neq q $ represents a coupling between
near-degenerate modes. Equation (\ref{calTsepterms}) shows that the
coupling occurs between modes either with same degree $l$ (but
generally different radial orders) or with degrees that differ by
$\pm 2$. The frequency correction, $\omega^{\rm T}$, for a single
mode can be obtained from the diagonal element $\mathcal{T}_{kk}$ as
\begin{equation}
\sigma^{\rm T}=\frac{\omega^{\rm T}}{\sqrt{GM/R^3}}
=\overline{\mathcal{T}}_{kk} \equiv\sigma_2^{\rm T}+\sigma_3^{\rm T}
\; ,
\end{equation}
with the second-order contribution
\begin{equation}
\begin{array}{rl}
\sigma_2^{\rm
T}=\frac{1}{2}(\frac{\displaystyle\sigma_0}{\displaystyle\bar{I}})
(\frac{\displaystyle\sigma_{\bar{\Omega}}}{\displaystyle\sigma_0})^2
\displaystyle\int_0^1  dx
\bar{\tilde{\rho}}x^4 &(\Lambda_{k+1}\mid\tau_{k+1}\mid^2\\
&~~~+\Lambda_{k-1}\mid\hat{\tau}_{k-1}\mid^2) \; , \\
\end{array}
\label{sigTsecondord}
\end{equation}
and the third-order contribution
\begin{equation}
\begin{array}{rl}
\sigma_3^{\rm T}=&m
(\frac{\displaystyle\sigma_0}{\displaystyle\bar{I}})
(\frac{\displaystyle\sigma_{\bar{\Omega}}}{\displaystyle\sigma_0})^3
\left\{ \frac{\displaystyle C_{\rm L}-1-J_1}{\displaystyle
2}\displaystyle\int_0^1  dx
\bar{\tilde{\rho}}x^4 \times \right.\\
&~~~~~\times (\Lambda_{k+1}\mid\tau_{k+1}\mid^2 +
\Lambda_{k-1}\mid\hat{\tau}_{k-1}\mid^2) \\
&\left. -\displaystyle\int_0^1dx \bar{\tilde{\rho}}x^4(1+\eta)
(\mid\tau_{k+1}\mid^2+\mid\hat{\tau}_{k-1}\mid^2)\right\} \; .\\
\end{array}
\label{sigTthirdord}
\end{equation}
Equations (\ref{calTsepterms}) to (\ref{sigTthirdord}) are identical
to Eqs. (B3)--(B7) of S98.

Note that to separate the frequency correction into the second and
third order in terms of $\epsilon=\sigma_{\bar{\Omega}}/\sigma_0$,
one also needs to use the following approximations:
\begin{equation}
\begin{array}{rl}
\frac{\displaystyle\omega_0}{\displaystyle\omega_0^{(0)}}=
\frac{\displaystyle\sigma_0}{\displaystyle\sigma_0^{(0)}}&\simeq
1+\left(1-\frac{\displaystyle\sigma_0^{(0)}}{\displaystyle\sigma_0}\right) \\
&=1+\frac{\displaystyle\sigma_{\bar{\Omega}}}{\displaystyle\sigma_0}
\left(\frac{\displaystyle\sigma_0-\sigma_0^{(0)}}
{\displaystyle\sigma_{\bar{\Omega}}} \right)\\
\noalign{\vskip 5pt}
&=1+m\frac{\displaystyle\bar{\Omega}}{\displaystyle\omega_0}(C_{\rm L}-1-J_1) \; ,\\
\end{array}
\end{equation}
\begin{equation}
\begin{array}{rl}
\frac{\displaystyle\sigma_{\bar{\Omega}}^2}{\displaystyle\bar{J}}
&\simeq \left( \frac{\displaystyle\sigma_0}{\displaystyle\bar{I}}
\right)
\left(\frac{\displaystyle\sigma_{\bar{\Omega}}}{\displaystyle\sigma_0}
\right)^2
\left[1+\frac{\displaystyle\sigma_{\bar{\Omega}}}{\displaystyle\sigma_0}
\left(\frac{\displaystyle\sigma_0-\sigma_0^{(0)}}
{\displaystyle\sigma_{\bar{\Omega}}} \right)
\right] \\
&\propto O(\bar{\Omega}^2)+O(\bar{\Omega}^3) \; , \\
\end{array}
\end{equation}
in which it is used that
\begin{equation}
\frac{\sigma_0-\sigma_0^{(0)}}{\sigma_{\bar{\Omega}}}=m(C_{\rm
L}-1-J_1)\propto
O\left({\frac{\bar{\Omega}}{\bar{\Omega}}}\right)=O(1) \; .
\end{equation}

\subsection{Non-spherically symmetric distortion: $\mathcal{D}_{kq}$}
Following again S98, the coefficients $\mathcal{D}_{kq}$ can be
obtained in an explicitly symmetric form as follows:
\begin{equation}
\begin{array}{rl}
\mathcal{D}_{kq}=&\mathcal{D}_{qk}=\frac{1}{\displaystyle 2J_{kq}}
\int d^3\mathbf{x}
\left\{(\Gamma_1p_2)\nabla\cdot\xi_{p0k}^{\ast}\nabla\cdot\xi_{p0q}+\right. \\
&~\rho_2(\xi_{p0q}\cdot\nabla\tilde{\phi}_k^{'\ast}+
\xi_{p0k}^{\ast}\cdot\nabla\tilde{\phi}_q^{'}) \\
&~-\frac{\displaystyle 1}{\displaystyle\tilde{\rho}}\nabla p_2\cdot
(\xi_{p0q}\tilde{\rho}_k^{'\ast}+
\xi_{p0k}^{\ast}\tilde{\rho}_q^{'})+
\tilde{\rho}r^2W \\
&\left.~-\rho_2\hat{\omega}_0^2\xi_{p0k}^{\ast}\cdot
\xi_{p0q}\right\} \; , \\
\end{array}
\label{symexpDinter}
\end{equation}
where
\begin{eqnarray}
W&=&y_ky_qY_k^{\ast}Y_q\left[P_2w_1+(1-P_2)
\frac{2}{3}r\frac{d\Omega^2}{dr}\right]
-\frac{1}{3}r\frac{d\Omega^2}{dr}\times\nonumber \\
&\times&(y_kz_qY_k^{\ast}\nabla_{\rm H} Y_q+y_qz_kY_q\nabla_{\rm H}
Y_k^{\ast}) \cdot\nabla_{\rm H} P_2 \; ,
\end{eqnarray}
where $Y_k \equiv Y_{l_k}^{m_k}$; here $P_2$ is the second Legendre
polynomial and $w_1$ is defined by
\begin{equation}
w_1=\frac{d}{dr}\left(\frac{1}{\tilde{\rho}}\right)\frac{d
p_{22}}{dr}+ \frac{1}{\tilde{\rho}}\frac{d}{dr}
\left(\frac{\rho_{22}}{\tilde{\rho}}\right)\frac{d\tilde{p}}{dr}
\label{w1} \; .
\end{equation}
The relations (\ref{symexpDinter}) to (\ref{w1}) are identical to
relations (B8) to (B10) in S98. Note that $\mathcal{D}_{kq}$ is
non-zero only when $m_k = m_q$.

The angular parts of the integral Eq.~(\ref{symexpDinter}) can be
evaluated analytically. Finally the result can be written in
dimensionless form as follows:
\begin{equation}
\begin{array}{rl}
\overline{\mathcal{D}}_{kq}&\equiv
\frac{\displaystyle\mathcal{D}_{kq}}{\displaystyle\sqrt{GM/R^3}} \\
&=\delta_{l_k l_q}\bar{D}+\delta_{l_k l_q+2}(
\frac{3}{2}\beta_{k}\beta_{q+1})\bar{D}_{kq}\\
&~~~~~~~~~~~+\delta_{l_k
l_q-2}(\frac{3}{2}\beta_{k+1}\beta_{q})\bar{D}_{kq}
\; , \\
\end{array}
\label{calDsepterms}
\end{equation}
with
\begin{equation}
\bar{D}=\mathcal{Q}_{kk2}\bar{D}_{kq}+
\frac{\sigma_{\bar{\Omega}}^2}{\bar{J}_{kq}}\int_0^1dx
\bar{\tilde{\rho}}x^4y_ky_qb_2 \; , \label{barDdef}
\end{equation}
and
\begin{eqnarray}
\bar{D}_{kq}=\bar{D}_{kq}^{(1)}+\bar{D}_{kq}^{(2)}+\bar{D}_{kq}^{(3)}+
\bar{D}_{kq}^{(4)}+\bar{D}_{kq}^{(5)} \; . \label{Dkq}
\end{eqnarray}
where the expressions for the different terms appearing in
$\bar{D}_{kq}$, are given in Appendix \ref{Appendix-Dkq}.

The diagonal elements $\overline{\mathcal{D}}_{kk}$ reduce to the
frequency correction $\omega^{\rm D}$ as
\begin{equation}
\sigma^{\rm D}=\frac{\omega^{\rm D}}{\sqrt{GM/R^3}}=
\overline{\mathcal{D}}_{kk}=\sigma_2^{\rm D}+\sigma_3^{\rm D} \; ,
\end{equation}
where the second-order contribution is
\begin{equation}
\sigma_2^{\rm D}= \left(\frac{\sigma_0}{\bar{I}}\right)
\left(\frac{\sigma_{\bar{\Omega}}}{\sigma_0}\right)^2
\left(\frac{\displaystyle\bar{J}_{kk}}
{\displaystyle\sigma_{\bar{\Omega}}^2}\right) \bar{D} \; ,
\label{sigDsecondord}
\end{equation}
and the third-order contribution is
\begin{equation}
\begin{array}{rl}
\sigma_3^{\rm D} &=
\left(\frac{\displaystyle\sigma_0}{\displaystyle\bar{I}}\right)
\left(\frac{\displaystyle\sigma_{\bar{\Omega}}}{\displaystyle\sigma_0}\right)^3
\Bigl(\frac{\displaystyle\sigma_0-\sigma_0^{(0)}}
{\displaystyle\sigma_{\bar{\Omega}}}\Bigr)
\left(\frac{\displaystyle\bar{J}_{kk}}
{\displaystyle\sigma_{\bar{\Omega}}^2}\right) \bar{D}\\
&=m(C_{\rm L}-1-J_1)\left(\frac{\displaystyle\sigma_0}
{\displaystyle\bar{I}}\right)
\left(\frac{\displaystyle\sigma_{\bar{\Omega}}}{\displaystyle\sigma_0}\right)^3
\left(\frac{\displaystyle\bar{J}_{kk}}
{\displaystyle\sigma_{\bar{\Omega}}^2}\right) \bar{D} \; , \\
\end{array}
\label{sigDthirdord}
\end{equation}
in which $\bar{D}$ is given by Eq.~(\ref{barDdef}).

Note that the relations (\ref{calDsepterms}) to (\ref{sigDthirdord})
differ from the relations (B12) to (B21) in S98. The differences
come from the fact that S98 have further performed integrations by
parts in order to remove $d\ln \tilde{\rho}/d\ln r=-(A+V_g)$ which
sometimes can cause numerical oscillations of the integrand of Eq.
(\ref{symexpDinter}). For the models which we have considered (see
Sect. 5) no such numerical problem exists and we choose to compute
the coefficient directly from the definition in
Eq.~(\ref{symexpDinter}). We furthermore note that the partial
integration gives rise to surface terms which are neglected by S98;
in fact, for models without an extended atmosphere they become
considerable. We return to this subject in
Sect.~\ref{Appendix-surface}.

Also in the third-order terms of the relations, the approximation $z
\simeq y_t/C\sigma_0^2$, which is identical with the non-rotating
case, instead of its exact relation Eq. (\ref{ampleqn1}) was used by
S98. We find in our numerical results that it is not true and
Eq.~(\ref{ampleqn1}) must be used (see Sect.
\ref{Appendix-approximation}).

\subsection{Distortion and Coriolis coupling: $\mathcal{C}_{kq}$}
As in S98, the coupling coefficients $\mathcal{C}_{kq}$ can be
obtained as:
\begin{equation}
\begin{array}{rl}
\overline{\mathcal{C}}_{kq}=\overline{\mathcal{C}}_{qk}
&\equiv\frac{\displaystyle\mathcal{C}_{kq}}{\displaystyle\sqrt{GM/R^3}} \\
&=\delta_{l_k l_q}\bar{C}+\delta_{l_k l_q+2}
(\frac{3}{2}\beta_{k}\beta_{q+1})\bar{C}_{kq} \\
&~~~~~~~~~~+\delta_{l_k
l_q-2}(\frac{3}{2}\beta_{k+1}\beta_{q})\bar{C}_{kq}
\; , \\
\end{array}
\label{calCsepterms}
\end{equation}
where
\begin{equation}
\begin{array}{rl}
\bar{C}=&\frac{\displaystyle m}{\displaystyle 2\bar{J}_{kq}}
\sigma_{\bar{\Omega}}^3(\sigma_{0k}+\sigma_{0q})\displaystyle\int_0^1
 dx
\bar{\tilde{\rho}}x^4C(1+\eta)\times\\
&\times[b_2-(A+V_g)u_2]\times\\
&\times[z_kz_q+\mathcal{Q}_{kq2}(y_kz_q+y_qz_k+3z_kz_q)] \; ,\\
\end{array}
\end{equation}
\begin{equation}
\begin{array}{rl}
\bar{C}_{kq}=&\frac{\displaystyle m}{\displaystyle 2\bar{J}_{kq}}
\sigma_{\bar{\Omega}}^3(\sigma_{0k}+\sigma_{0q})\displaystyle\int_0^1
 dx
\bar{\tilde{\rho}}x^4C(1+\eta)\times\\
&\times[b_2-(A+V_g)u_2][y_kz_q+y_qz_k+3z_kz_q] \; ;\\
\end{array}
\label{Ckq}
\end{equation}
an expression for $\mathcal{Q}_{kq2} \equiv \int \sin \theta d\theta
d\phi Y_k^\ast Y_q P_2$ is given by Eq. (B18) of S98. The diagonal
coefficients correspond to the contribution to the frequency
correction $\omega^{\rm C}=\mathcal{C}_{kk}$. This contribution has
only terms of third order as follows:
\begin{equation}
\begin{array}{rl}
\sigma^{\rm C}&=\overline{\mathcal{C}}_{kk}\equiv
\frac{\displaystyle \omega^{\rm C}}{\displaystyle \sqrt{GM/R^3}}
=\sigma_3^{\rm C} \\
\noalign{\vskip 3pt} &=\frac{\displaystyle m}{\displaystyle\bar{I}}
\sigma_{\bar{\Omega}}^3\displaystyle\int_0^1  dx
\bar{\tilde{\rho}}x^4C(1+\eta)[b_2-(A+V_g)u_2]\times\\
&~~~~~~~~\times[z^2+\mathcal{Q}_{kk2}(2yz+3z^2)] \; .\\
\end{array}
\label{sigCthirdord}
\end{equation}
Equations (\ref{calCsepterms}) to (\ref{Ckq}) are identical to the
relation (B22) in S98. However, the relation (\ref{sigCthirdord})
differs from Eq. (B25) in S98 where again a partial integration has
been carried out to eliminate the density derivative.

Note that Eqs. (\ref{sigTthirdord}), (\ref{sigDthirdord}) and
(\ref{sigCthirdord}) show that for the case $m=0$ there are no
third-order contributions to the frequency corrections.

The expressions for the separate terms in $\mathcal{H}_{kq}$ (see
Eq.~(\ref{calHmatdef})), i.e., Eqs. (\ref{calTsepterms}),
(\ref{symexpDinter}), and (\ref{calCsepterms}), are all explicitly
symmetric and real-valued. Therefore one can conclude that the total
coupling coefficient is symmetric, real-valued and therefore also
hermitian, $\mathcal{H}_{kq}=\mathcal{H}_{qk}$.


\subsection{Close frequencies and mode coupling}
The existence of close frequencies can lead to large values of the
perturbation terms to the extent that it invalidates the
perturbative approach outlined in the previous sections. For such
cases one needs to use a degenerate perturbation formalism.
Following S98, the zeroth-order eigenfunction of an individual mode
is written as a superposition of the degenerate eigenfunctions:
\begin{equation}
\xi=\sum_k\mathcal{A}_k\xi_{0k} \; , \label{degeneigfun}
\end{equation}
where $\xi_{0k}$ is the degenerate eigenfunction of the zeroth-order
system and the amplitudes $\mathcal{A}_k$ are expansion coefficients
which can be normalized to 1. As same in S98, the solution of the
eigenvalue problem for the eigenfunction $\xi$, which corresponds to
an eigenvalue $\omega$, satisfies the following linear system:
\begin{equation}
\mathcal{A}_k\mu_{kk}+\sum_{q\neq k}\mathcal{A}_q\mu_{kq}=0 \; ,
\label{deglineq}
\end{equation}
where the components $\mu_{kq}$ are defined by
\begin{equation}
\begin{array}{rl}
\mu_{kk}&\equiv(\omega_k-\omega)2\omega_{0k}^{(0)}I_k \; , \\
\mu_{kq}&\equiv 2J_{kq}\mathcal{H}_{kq} \; ,~~~~~~{\rm for}~q\neq k \; . \\
\end{array}
\label{mufacdef}
\end{equation}
One notes that $\mu_{kq}=\mu_{qk}$ because of the symmetric
properties of $\mathcal{H}_{kq}$. The eigenvalues, $\omega$, can be
obtained from the existence condition of nontrivial solutions of Eq.
(\ref{deglineq}), i.e.,
\begin{equation}
\mathrm{det}[\mu_{ij}]=0 \; ,~~~~~~i,j=1, \ldots ,N \; ,
\label{zerodetlineq}
\end{equation}
where $\mathrm{det}$ is the determinant and $N$ is the total number
of the near-degenerate modes.

Here we consider the coupling between $N=2$ modes. Then Eq.
(\ref{zerodetlineq}) reduces to
\begin{equation}
(\omega_1-\omega)(\omega_2-\omega)-\mathcal{H}_{12}^2=0 \; .
\end{equation}
In this case the eigenfrequencies, $\omega=\omega_{\pm}$, can be
obtained trivially from the zeroth-order near-degenerate
eigenfrequencies and the total corrections due to rotation and
coupling are
\begin{equation}
\omega_{\pm}=\left(\frac{\omega_1+\omega_2}{2}\right)\pm
\sqrt{\left(\frac{\omega_1-\omega_2}{2}\right)^2
+\mathcal{H}_{12}^2}+O(\epsilon^4) \; . \label{omegaplusminus}
\end{equation}
Also, the normalized amplitude $\mathbf{\mathcal{A}}\equiv
(\mathcal{A}_1,\mathcal{A}_2)$ appearing in Eq.~(\ref{degeneigfun})
can be derived from Eq.~(\ref{deglineq}) as follows
\begin{equation}
\begin{array}{rl}
\mathcal{A}_1^{(\pm)}&=[1+(\mu_{11}^{(\pm)}/\mu_{12})^2]^{-1/2} \\
&=[1+(\mu_{12}/\mu_{22}^{(\pm)})^2]^{-1/2} \\
\mathcal{A}_2^{(\pm)}&=-(\mu_{11}^{(\pm)}/\mu_{12})\mathcal{A}_1^{(\pm)} \\
&=-(\mu_{12}/\mu_{22}^{(\pm)})\mathcal{A}_1^{(\pm)} \; , \\
\end{array}
\label{A1A2plusminus}
\end{equation}
where the quantities labeled by $(\pm)$ are related to the
eigenvalues $\omega_{\pm}$.

The coupling coefficients $\mathcal{H}_{kq}$ appearing in
Eq.~(\ref{mufacdef}) show that the near-degenerate coupling only
occurs between modes with the same $m$ and with either the same
degree $l$ (and different radial orders) or with degrees which
differ by $\pm2$.
\clearpage
\section{Oscillations of a rapidly rotating B star}\label{NR}
In order to calculate the effect of rotation on normal modes, I
consider a uniformly rotating, 12~$M_\odot$, ZAMS model generated by
the evolution code of Christensen-Dalsgaard (1982) (see also
Christensen-Dalsgaard \& Thompson 1999). The parameters of the model
are listed at Table \ref{Model-ZAMS}.

\begin{table}
\caption[]{Stellar parameters of a rotating zero-age main-sequence
star in solar units. $M$, $M_{\rm conv}$, $R$, $R_{\rm conv}$,
$p_{\rm c}$, and $\rho_{\rm c}$ are the total mass, the mass of
convective core, the radius, the radius of convective core, the
central pressure and density, and $\odot$ denotes solar values;
$\sigma_{\bar{\Omega}}$ and $\epsilon$ are the dimensionless mean
angular velocity and the perturbational expansion coefficient;
$T_{\rm dyn}$, $T_{\rm rot}$, and $V_{\rm rot}$ are the dynamical
time scale (free fall time), the equatorial period and velocity,
respectively. For comparison note that $T_{\odot \rm dyn}=0.5~ \rm
h$, $T_{\odot \rm rot}=25~\rm d$, $V_{\odot \rm rot}=2~\rm km \,
s^{-1}$.}
\begin{tabular}{ccc} \hline\hline
$M=12~M_\odot$                               & $M_{\rm conv}= 0.34~M$ &\\
$R=4.32~R_\odot$                             & $R_{\rm conv}=0.25~R$  &\\
$p_{\rm c}=3.72\times 10^{-1}~p_{\odot \rm c}$   & $\rho_{\rm c}=4.83\times 10^{-2}~\rho_{\odot \rm c}$ &\\
$\sigma_{\bar{\Omega}}=1.38\times10^{-1}$       & $\epsilon=\Omega/\omega=\sigma_{\bar{\Omega}}/2 \pi=2.2\times 10^{-2}$ &\\
$T_{\rm dyn}=\sqrt{R^3/GM}=1.15~\rm h$       & $T_{\rm rot}=2\pi R/V_{\rm rot}=2.15~\rm d$&\\
$V_{\rm rot}=R\Omega=R\sigma_{\bar{\Omega}}/T_{\rm dyn}=100~\rm km \, s^{-1}$ & &\\
\hline
\end{tabular}\\
\label{Model-ZAMS}
\end{table}

\begin{figure}
 \includegraphics{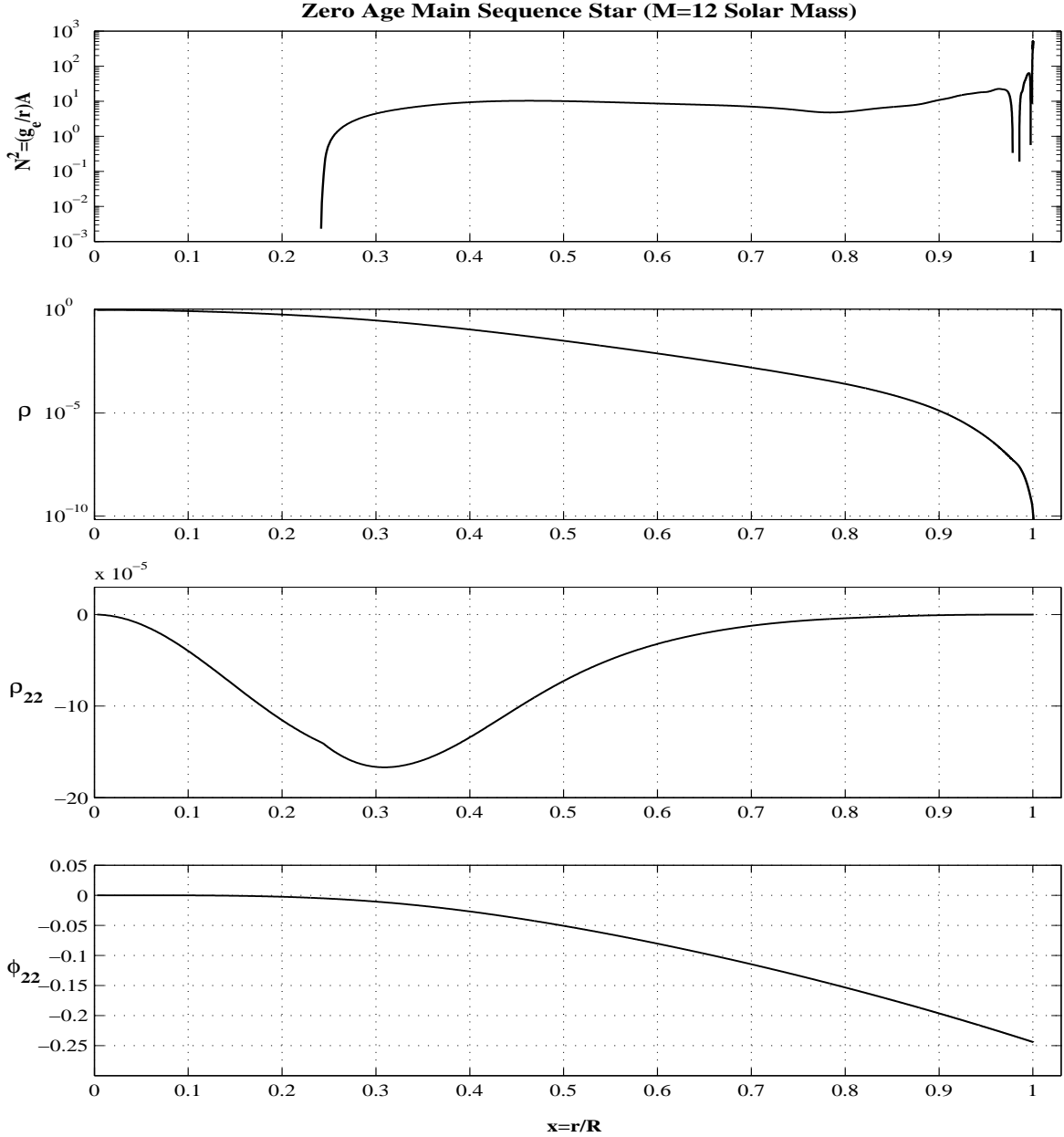}
      \vspace{19.6cm}
      \caption[]{Dimensionless equilibrium quantities including squared
      buoyancy frequency $N^2=(g_{\rm e}/r)A$
      with $A=(1/\Gamma_1)(d\ln p/d \ln x)-(d\ln \rho/d \ln x)$,
      spherically and non-spherically-symmetric contributions
      $\rho$ and $\rho_{22}$ to the density, and
      non-spherically-symmetric gravitational-potential contribution
      $\phi_{22}$, in units of $GM/R^3$, $\rho_{\rm c}$ and
      $R^2\bar{\Omega}^2$, respectively, against fractional radius
      $x=r/R$ for a zero-age main-sequence star with $M=12M_{\odot}$.}
         \label{N2-rho-rho22-phi22-ZAMS}
   \end{figure}

The behavior of some of equilibrium quantities of the model against
fractional radius, $x=r/R$ are shown in
Fig.~\ref{N2-rho-rho22-phi22-ZAMS}; It shows that: 1) From the
centre to the radius $x=0.25$, the star is in a convective regime
where $N^2<0$ and, outside of this radius is in a radiative regime
where $N^2>0$; 2) The spherically symmetric density $\rho$ decreases
smoothly from its maximum value to nearly zero at the surface. 3)
The non-spherically-symmetric correction to the density $\rho_{22}$
decreases to a negative maximum value at $x=0.3$ and then slowly
increases to zero at $x\simeq 0.9$; 4) The non-spherically-symmetric
correction to the gravitational potential $\phi_{22}$ is obtained as
the numerical solution of the Poisson relation, Eq. (12), by a
Runge-Kutta method with an adaptive step-size control. The absolute
value of $\phi_{22}$ increases smoothly to its maximum value at the
surface.

\subsection{Eigenfunctions}
The zero-order eigenfunctions are computed from the zero order
eigenvalue problem with the pulsation code of Christensen-Dalsgaard
(see Christensen-Dalsgaard \& Berthomieu 1991), modified according
to Eqs.~(\ref{ydrdifeq}) to (\ref{ampleqn1}).

\begin{figure}
 \includegraphics{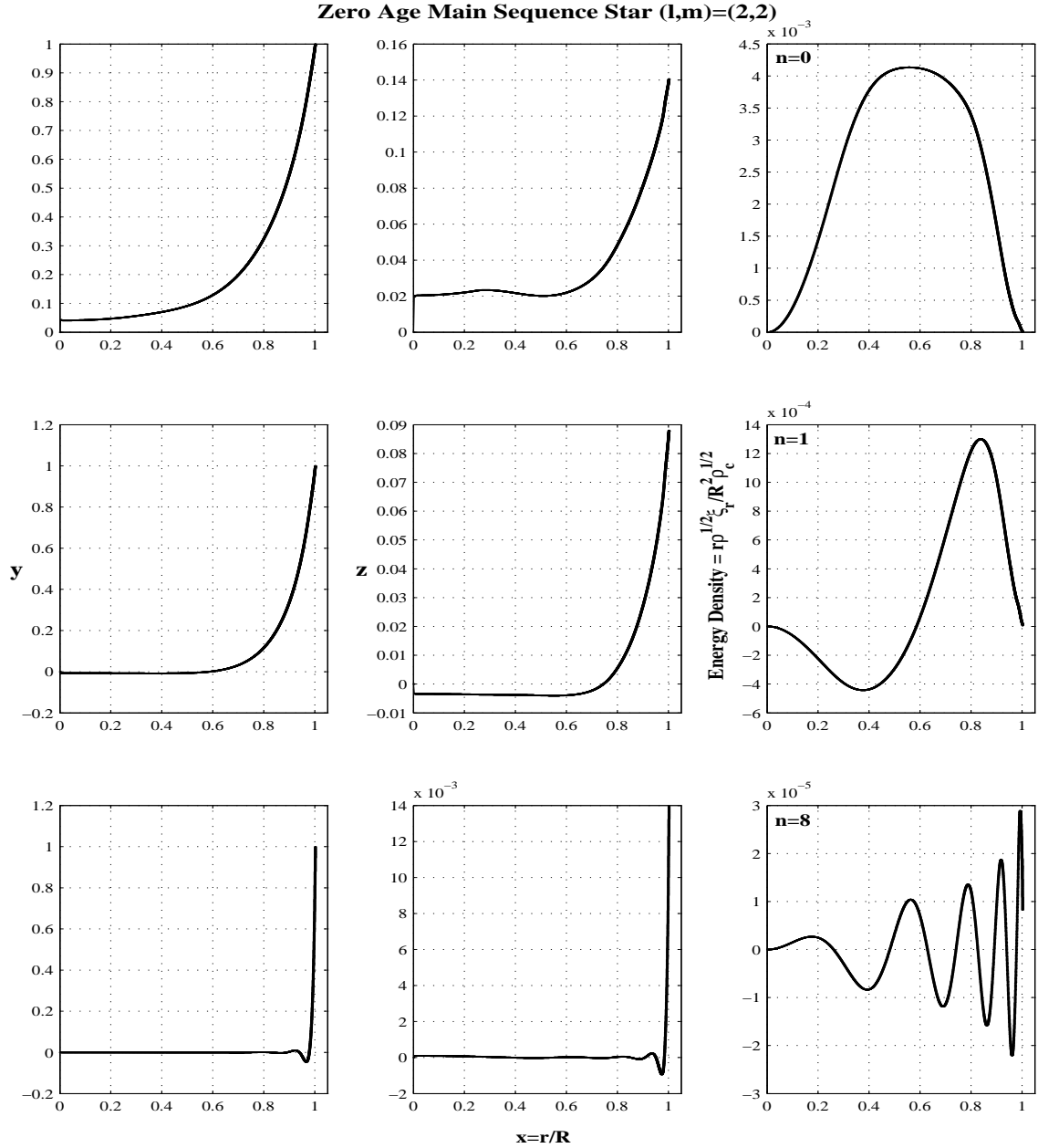}
      \vspace{19.6cm}
      \caption[]{Zero-order radial $y$
      (left) and horizontal $z$ components (middle) as well as
      $r\rho^{1/2}\xi_r/(R^2\rho_{\rm c}^{1/2})$, related to the
      energy density (right),
      against fractional radius $x=r/R$ for selected f and p modes
      with $(l,m)=(2,2)$ and $n=(0,1,8)$ for the model
      described in Table \ref{Model-ZAMS}.
      }
         \label{ZAMS-yzED-l2m2}
   \end{figure}

In Fig. \ref{ZAMS-yzED-l2m2}, the radial ($y$) and horizontal ($z$)
components of the zero-order poloidal eigenfunctions as well as
$r\rho^{1/2}\xi_r/(R^2\rho_{\rm c}^{1/2})$ related to the radial
energy density, where $\xi_r=ry$ is the radial displacement, are
plotted against the fractional radius $x=r/R$ for the selected f and
p modes with $(l,m)=(2,2)$ and $n=(0,1,8)$. Note that in the case of
radial oscillation ($l=0$), the results are derived from the reduced
set of differential equations composed of Eqs. (\ref{ydrdifeqrad})
and (\ref{ytdrdifeqrad}).

Figure \ref{ZAMS-yzED-l2m2} shows that: 1) The number of nodes
coincides with the radial order $n$. 2) At lower radial order, $n$,
the energy is distributed throughout most of the star. At higher
radial order, $n$, the displacement is large only in the outer part
of the model. 3) The magnitude of the horizontal component $z$
decreases with increasing $n$ (see also
Sobouti (1980)). In other words, the higher-order p modes oscillate
radially rather than horizontally because they are acoustic and
longitudinal waves.

\subsection{Eigenfrequencies and corrections}
The zero-order eigenfrequency, $\sigma_0$, is derived from numerical
solutions of Eqs.~(\ref{ydrdifeq}) to (\ref{ampleqn1}) by the
modified pulsation code;
note that using the eigensystem in Eqs.~(\ref{ydrdifeq}) to
(\ref{ampleqn1}) the first-order frequency correction, $\sigma_1$,
is implicitly included in $\sigma_0$ (see Eqs.~(\ref{ome00exp}) and
(\ref{ome1def})). The second- and third-order Coriolis
contributions, ($\sigma_2^{\rm T},~ \sigma_3^{\rm T}$), the second-
and third-order non-spherically-symmetric distortions,
($\sigma_2^{\rm D},~ \sigma_3^{\rm D}$), and the third-order
distortion and Coriolis coupling, $\sigma_3^{\rm C}$, are derived
from numerical integrations of Eqs. (\ref{sigTsecondord}),
(\ref{sigTthirdord}), (\ref{sigDsecondord}), (\ref{sigDthirdord}),
and (\ref{sigCthirdord}).

In Tables \ref{Table-Sigma0-T23-D23-C-total-l1m1-ZAMS} to
\ref{Table-Sigma0-T23-D23-C-total-l2m2-ZAMS} the results of
different contributions of frequency corrections due to effect of
rotation up to third order are tabulated. In each table the selected
p modes with $(l,m)=(0,0)$, $(1,1)$, $(2,1)$ and $(2,2)$ with $n=(0,
\ldots ,14)$, $(1, \ldots ,15)$, and $(0, \ldots ,14)$ respectively,
are considered. The fundamental modes are known with $n=0$ and
labelled by f. The modes with $n\geq 1$ are labelled by ${\rm p}_1,
\ldots ,{\rm p}_n$.

Tables \ref{Table-Sigma0-T23-D23-C-total-l1m1-ZAMS} to
\ref{Table-Sigma0-T23-D23-C-total-l2m2-ZAMS} show that: 1) The
values of zero-order eigenfrequency, $\sigma_0$, increase when the
radial order $n$ increases. 2) For the case $m=0$, the odd-order
frequency corrections, i.e., in the present case those of first and
third order, vanish. In this case only the second-order frequency
corrections exist (see Eqs. (\ref{ome1calc}), (\ref{sigTsecondord}),
(\ref{sigTthirdord}), (\ref{sigDsecondord}), (\ref{sigDthirdord}),
and (\ref{sigCthirdord})). 3) For radial oscillation ($l=0$) only
the second-order Coriolis contribution exists and the second-order
non-spherically-symmetric distortion which results from centrifugal
force vanishes for a uniform rotation (see
Eq.~(\ref{sigDsecondord})). 4) In non-radial cases, at lower radial
orders, the Coriolis contribution and non-spherically-symmetric
distortion have the same order of magnitude. But at higher orders,
the magnitude of the non-spherically-symmetric distortion becomes
greater than Coriolis contribution. 5) With increasing $n$, the
frequency correction due to distortion and Coriolis coupling
increases and decreases alternatively.

\begin{figure}
 \includegraphics{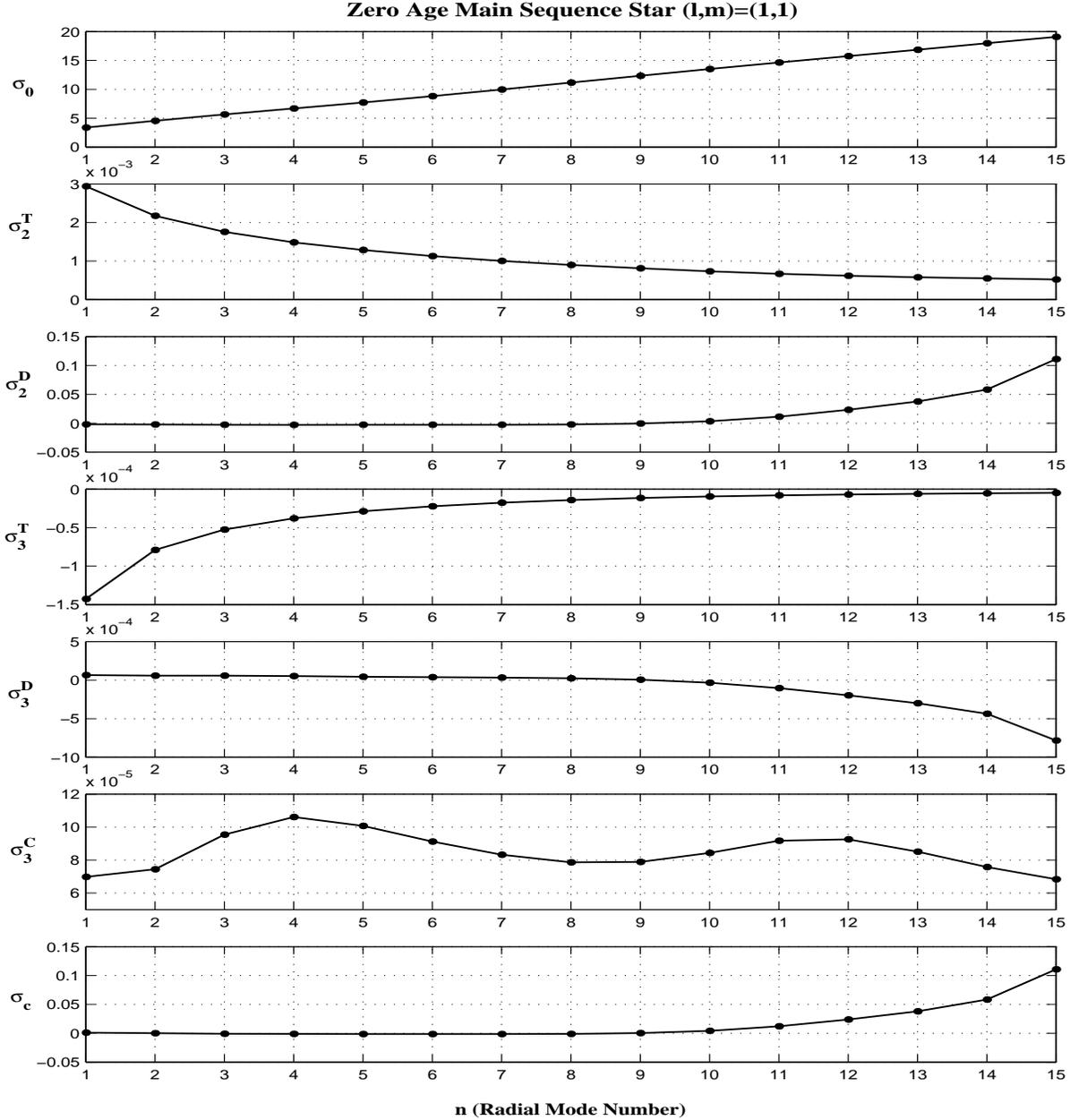}
      \vspace{19.6cm}
      \caption[]{Frequency corrections against radial order
      $n$ for p modes with $(l,m)=(1,1)$ and $n=(1, \ldots ,15)$
      for a zero-age main-sequence star with $M=12M_{\odot}$.
      Here $\sigma_0$ is the zero-order eigenfrequency,
$\sigma_2^{\rm T}$ and $\sigma_3^{\rm T}$ are the second
       and the third-order Coriolis contributions,
       and $\sigma_2^{\rm D}$ and $\sigma_3^{\rm D}$ are the second-
       and the third-order
       non-spherically-symmetric distortions,
       $\sigma_3^{\rm C}$ is the third-order
       distortion and Coriolis coupling,
       and $\sigma_{\rm c}=\sigma_2^{\rm T}+\sigma_3^{\rm T}
        +\sigma_2^{\rm D}+\sigma_3^{\rm D}+\sigma_3^{\rm C}$
is the total third-order frequency correction. All frequencies are
in units of $\sqrt{GM/R^3}=2.42\times 10^{-4}\, {\rm s}^{-1}$.
              }
         \label{Fig-Sigma0-T23-D23-C-H-l1m1-ZAMS}
   \end{figure}

\begin{figure}
 \includegraphics{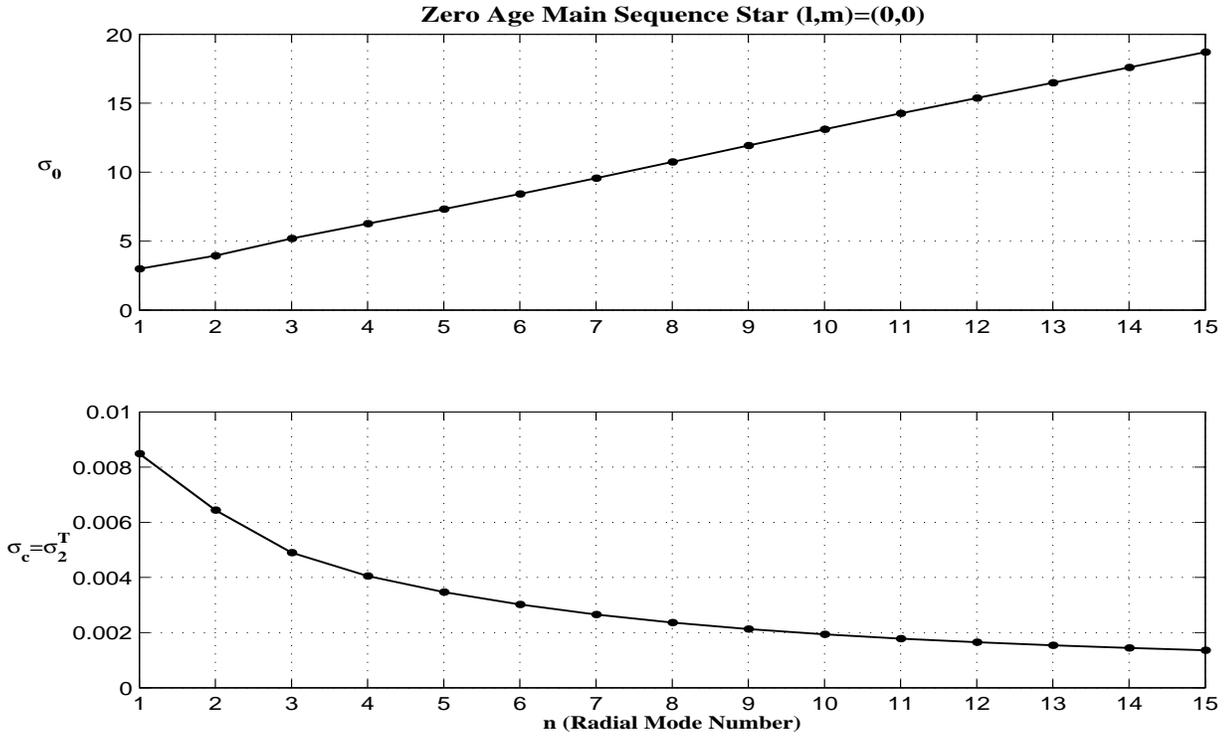}
      \vspace{16.5cm}
      \caption[]{Same as Fig.~\ref{Fig-Sigma0-T23-D23-C-H-l1m1-ZAMS},
for p modes with $(l,m)=(0,0)$ and $n=(1, \ldots ,15)$. There is no
$\sigma_3^{\rm T}$, $\sigma_3^{\rm D}$, and $\sigma_3^{\rm C}=0$ for
$m=0$ (see Eqs.~(\ref{sigTthirdord}), (\ref{sigDthirdord}), and
(\ref{sigCthirdord})). Also $\sigma_2^{\rm D}$ vanishes since
$l=m=0$ and the model is a uniformly rotating star (see
Eq.~(\ref{sigDsecondord})).}

  \label{Fig-Sigma0-T23-D23-C-H-l0m0-ZAMS}
   \end{figure}

\begin{figure}
 \includegraphics{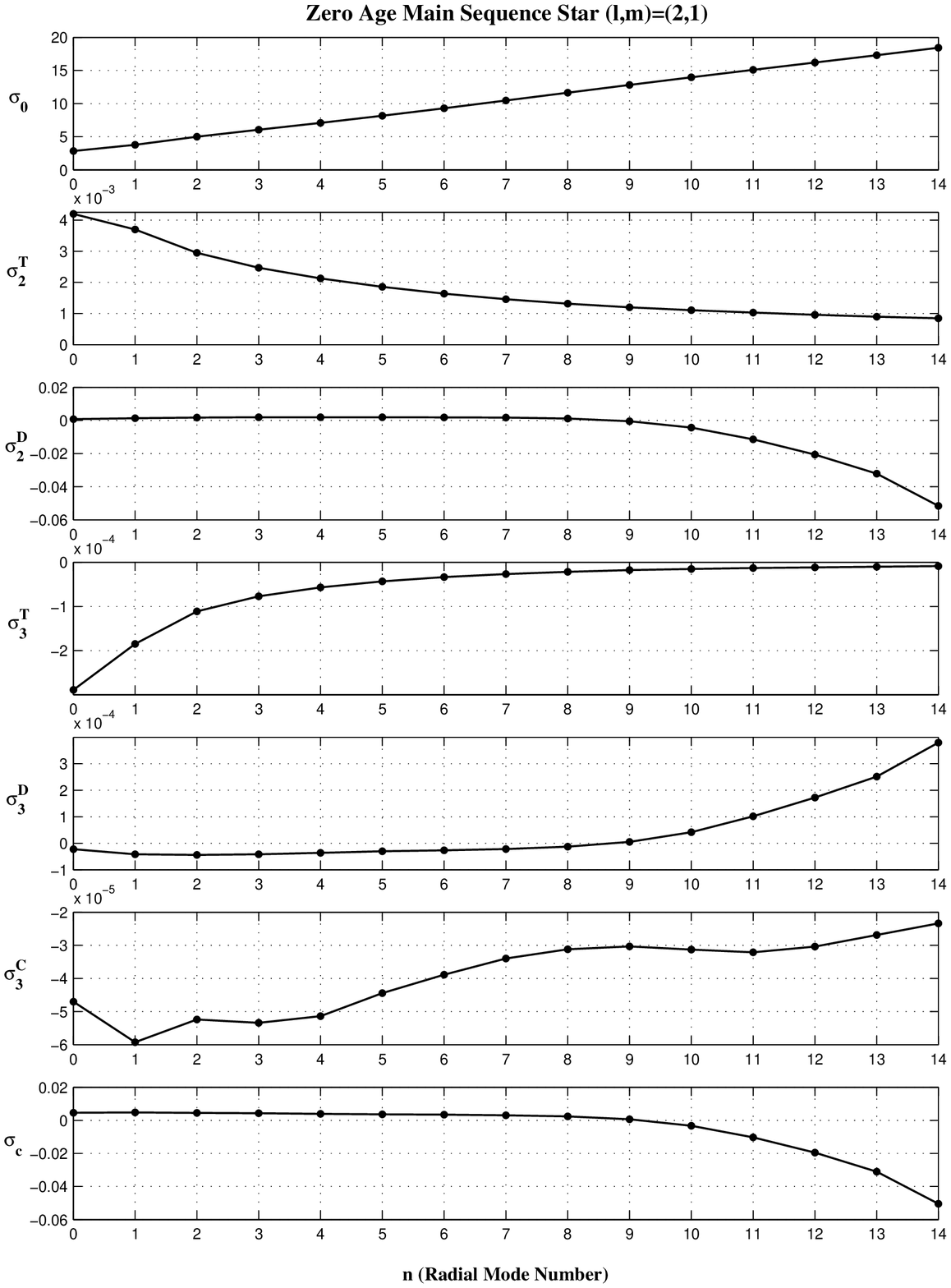}
      \vspace{19.6cm}
      \caption[]{Same as Fig.~\ref{Fig-Sigma0-T23-D23-C-H-l1m1-ZAMS}, for f and p modes with $(l,m)=(2,1)$ and
$n=(0, \ldots ,14)$.
              }
         \label{Fig-Sigma0-T23-D23-C-H-l2m1-ZAMS}
   \end{figure}

\begin{figure}
 \includegraphics{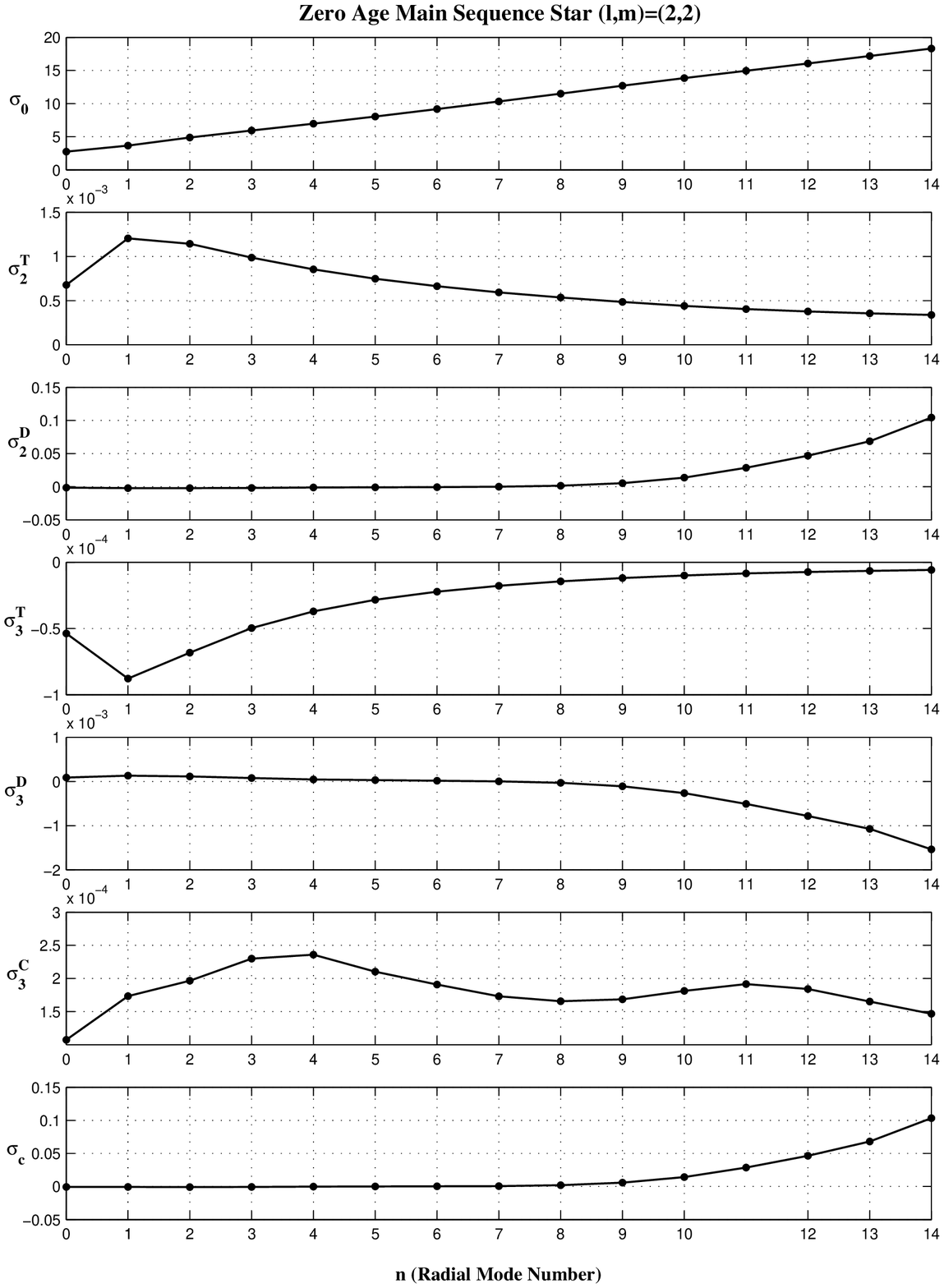}
      \vspace{19.6cm}
      \caption[]{Same as Fig.~\ref{Fig-Sigma0-T23-D23-C-H-l1m1-ZAMS}, for f and p modes with $(l,m)=(2,2)$ and
$n=(0, \ldots ,14)$.
              }
         \label{Fig-Sigma0-T23-D23-C-H-l2m2-ZAMS}
   \end{figure}

In Figs. \ref{Fig-Sigma0-T23-D23-C-H-l1m1-ZAMS} to
\ref{Fig-Sigma0-T23-D23-C-H-l2m2-ZAMS} the results in Tables
\ref{Table-Sigma0-T23-D23-C-total-l1m1-ZAMS} to
\ref{Table-Sigma0-T23-D23-C-total-l2m2-ZAMS} are plotted. The plots
show that for high-order p modes:
1) $\sigma_0$ increases as a linear function of $n$, which is very
similar to the pattern of vibration in a simple string; this also
follows from asymptotic analyses of high-order acoustic modes (e.g.,
Vandakurov 1967; Tassoul 1980). 2) ($\sigma_2^{\rm T}$,
$\sigma_2^{\rm D}$) and ($\sigma_3^{\rm T}$, $\sigma_3^{\rm D}$)
have almost regular asymptotic behaviors as $1/\sigma_0\propto1/n$
and $1/\sigma_0^2\propto1/n^2$, respectively. See Eqs.
(\ref{sigTsecondord}), (\ref{sigDsecondord}) and
(\ref{sigTthirdord}), (\ref{sigDthirdord}). 4) The distortion and
Coriolis coupling, $\sigma_3^{\rm C}$, has no clear asymptotic
relation (see Eq. (\ref{sigCthirdord})). 5) The total frequency
shift corrected up to third order, $\sigma_{\rm c}$, for radial and
non-radial modes decreases and increases, respectively, with
increasing $n$; in the non-radial case this is caused by the
dominance of the term $\sigma_2^{\rm D}$ which vanishes in the
radial case. Note that for the case of $(l,m)=(2,2)$, in the
diagrams of $\sigma_2^{\rm T}$ and $\sigma_3^{\rm T}$ a break
between f and p$_1$ appears. This happens because the inertia for
the f mode is two order of magnitude greater than for ${\rm p}_1$.

In Table \ref{TwoModesCoupling-ZAMS}, the results on third-order
frequency corrections for the case of two near-degenerate modes,
derived from Eqs. (\ref{omegaplusminus})--(\ref{A1A2plusminus}), are
tabulated. As discussed above the coupling exists only for the two
near-degenerate poloidal modes belonging to the same $m$ but with
$l$ differing by $\pm 2$.
\clearpage
\clearpage
\clearpage
\begin{table}
\caption[]{Values of eigenfrequency $\sigma_0$ (including the
$O(\Omega)$ contribution from rotation), second- and third-order
Coriolis contributions $\sigma_2^{\rm T}$ and $\sigma_3^{\rm T}$,
second- and third-order non-spherically-symmetric distortions
$\sigma_2^{\rm D}$ and $\sigma_3^{\rm D}$, third-order distortion
and Coriolis coupling $\sigma_3^{\rm C}$, and total frequency
$\sigma_{\rm tot}= \sigma_0+\sigma_{\rm c}$ corrected up to third
order, for p modes in a zero-age main-sequence star with
$M=12M_{\odot}$, $(l,m)=(1,1)$ and $n=(1, \ldots ,15)$. Here
$\sigma_{\rm c}=\sigma_2^{\rm T}+\sigma_3^{\rm T}+\sigma_2^{\rm D}
+\sigma_3^{\rm D}+\sigma_3^{\rm C}$ is the total third-order
frequency correction. All frequencies are in units of
$\sqrt{GM/R^3}=2.42\times 10^{-4}\, {\rm s}^{-1}$. }
\begin{center}
\begin{tabular}{rrrcrcrcrcccccc} \hline
\multicolumn{1}{c}{$Mode$}&$n$&\multicolumn{1}{c}{$\sigma_0$}&$\sigma_2^{\rm
T}$&\multicolumn{1}{c}{$\sigma_2^{\rm D}$}&$\sigma_3^{\rm
T}$&\multicolumn{1}{c}{$\sigma_3^{\rm D}$}&$\sigma_3^{\rm
C}$&\multicolumn{1}{c}{$\sigma_{\rm tot}$}&\\\hline
  ${\rm p}_1$&1&   3.3700 &   2.940$\times 10^{-3}$&    $-1.873\times 10^{-3}$&  $-1.43\times 10^{-4}$& $6.53\times 10^{-5}$&   6.98$\times 10^{-5}$&   3.3711 &\\
  ${\rm p}_2$& 2&   4.5499 &  2.173$\times 10^{-3}$&    $-2.227\times 10^{-3}$&  $-7.91\times 10^{-5}$& $5.85\times 10^{-5}$&   7.45$\times 10^{-5}$&   4.5499 &\\
  ${\rm p}_3$&3&   5.6388 &   1.759$\times 10^{-3}$&    $-2.695\times 10^{-3}$&  $-5.27\times 10^{-5}$& $5.87\times 10^{-5}$&   9.54$\times 10^{-5}$&   5.6380 &\\
  ${\rm p}_4$& 4&   6.6669 &  1.483$\times 10^{-3}$&   $-2.777\times 10^{-3}$&   $-3.81\times 10^{-5}$& $5.22\times 10^{-5}$&   1.06$\times 10^{-4}$&   6.6657 &\\
  ${\rm p}_5$& 5&   7.7061 &  1.285$\times 10^{-3}$&    $-2.701\times 10^{-3}$&  $-2.88\times 10^{-5}$&  $4.44\times 10^{-5}$&   1.01$\times 10^{-4}$&   7.7048 &\\
  ${\rm p}_6$& 6&   8.8212 &  1.129$\times 10^{-3}$&   $-2.648\times 10^{-3}$&   $-2.23\times 10^{-5}$&   $3.85\times 10^{-5}$&   9.12$\times 10^{-5}$&   8.8198 &\\
  ${\rm p}_7$& 7&   9.9704 &  1.003$\times 10^{-3}$&   $-2.522\times 10^{-3}$&   $-1.77\times 10^{-5}$&   $3.28\times 10^{-5}$&   8.32$\times 10^{-5}$&   9.9690 &\\
  ${\rm p}_8$& 8&   11.152  &  8.986$\times 10^{-4}$&    $-2.051\times 10^{-3}$&  $-1.42\times 10^{-5}$&  $2.40\times 10^{-5}$&   7.87$\times 10^{-5}$&   11.151 &\\
  ${\rm p}_9$& 9&   12.336 &  8.098$\times 10^{-4}$&   $-5.514\times 10^{-4}$&   $-1.17\times 10^{-5}$&  $5.88\times 10^{-6}$&   7.89$\times 10^{-5}$&   12.336 &\\
  ${\rm p}_{10}$&10&   13.503 &  7.323$\times 10^{-4}$&   $3.330\times 10^{-3}$& $-9.67\times 10^{-6}$& $-3.26\times 10^{-5}$&   8.44$\times 10^{-5}$&   13.507 &\\
  ${\rm p}_{11}$&11&   14.635 &  6.661$\times 10^{-4}$&   $1.135\times 10^{-2}$& $-8.15\times 10^{-6}$& $-1.03\times 10^{-4}$&   9.17$\times 10^{-5}$&   14.647 &\\
  ${\rm p}_{12}$&12&   15.741 &  6.153$\times 10^{-4}$&   $2.329\times 10^{-2}$& $-7.01\times 10^{-6}$& $-1.97\times 10^{-4}$&   9.26$\times 10^{-5}$&   15.765 &\\
  ${\rm p}_{13}$& 13&   16.851 & 5.780$\times 10^{-4}$&   $3.771\times 10^{-2}$& $-6.17\times 10^{-6}$& $-2.99\times 10^{-4}$&   8.51$\times 10^{-5}$&   16.889 &\\
  ${\rm p}_{14}$&14&   17.971 &  5.472$\times 10^{-4}$&   $5.833\times 10^{-2}$& $-5.48\times 10^{-6}$&  $-4.35\times 10^{-4}$&   7.58$\times 10^{-5}$&   18.029 &\\
  ${\rm p}_{15}$&15&   19.065 &  5.197$\times 10^{-4}$&   $1.110\times 10^{-1}$& $-4.92\times 10^{-6}$&  $-7.83\times 10^{-4}$&   6.84$\times 10^{-5}$&   19.176 &\\
\hline
\end{tabular}\\
\end{center}
\label{Table-Sigma0-T23-D23-C-total-l1m1-ZAMS}
\end{table}
\clearpage
\begin{table}
\caption[]{Same as Table
\ref{Table-Sigma0-T23-D23-C-total-l1m1-ZAMS}, for p modes with
$(l,m)=(0,0)$ and $n=(1, \ldots ,15)$. There is no $\sigma_3^{\rm
T}$, $\sigma_3^{\rm D}$, and $\sigma_3^{\rm C}=0$ for $m=0$ (see
Eqs. (53), (65), and (69)). Also $\sigma_2^{\rm D}$ vanishes since
$l=m=0$ and the model is a uniformly rotating star (see Eq.~(64)).}
\begin{center}
\begin{tabular}{rrrcrccccccccc} \hline
\multicolumn{1}{c}{$Mode$}&$n$&\multicolumn{1}{c}{$\sigma_0$}&$\sigma_2^{\rm
T}$&\multicolumn{1}{c}{$\sigma_{\rm tot}$}&\\\hline
 ${\rm p}_1$&1&   2.9941 &   8.488$\times 10^{-3}$&   3.0026 &\\
 ${\rm p}_2$& 2&   3.9460 &  6.440$\times 10^{-3}$&   3.9524 &\\
 ${\rm p}_3$& 3&   5.1853 &  4.901$\times 10^{-3}$&   5.1902 &\\
 ${\rm p}_4$& 4&   6.2697 &  4.053$\times 10^{-3}$&   6.2738 &\\
 ${\rm p}_5$& 5&   7.3206 &  3.471$\times 10^{-3}$&   7.3241 &\\
 ${\rm p}_6$& 6&   8.4203 &  3.018$\times 10^{-3}$&   8.4233 &\\
 ${\rm p}_7$&7&   9.5645 &   2.657$\times 10^{-3}$&   9.5671 &\\
 ${\rm p}_8$& 8&   10.742 &  2.366$\times 10^{-3}$&   10.745 &\\
 ${\rm p}_9$& 9&   11.932 &  2.130$\times 10^{-3}$&   11.934 &\\
${\rm p}_{10}$& 10&   13.112 &  1.938$\times 10^{-3}$&   13.114 &\\
${\rm p}_{11}$&11&   14.261 &   1.782$\times 10^{-3}$&   14.262 &\\
${\rm p}_{12}$& 12&   15.374 &  1.653$\times 10^{-3}$&   15.376 &\\
${\rm p}_{13}$& 13&   16.478 &  1.542$\times 10^{-3}$&   16.479 &\\
${\rm p}_{14}$& 14&   17.591 &  1.445$\times 10^{-3}$&   17.592 &\\
${\rm p}_{15}$& 15&   18.702 &  1.359$\times 10^{-3}$&   18.704 &\\
\hline
\end{tabular}\\
\end{center}
\label{Table-Sigma0-T23-D23-C-total-l0m0-ZAMS}
\end{table}
\begin{table}
\caption[]{Same as Table
\ref{Table-Sigma0-T23-D23-C-total-l1m1-ZAMS}, for f and p modes with
$(l,m)=(2,1)$ and $n=(0, \ldots ,14)$.}
\begin{center}
\begin{tabular}{rrrcrcrcrcccc} \hline
\multicolumn{1}{c}{$Mode$}&$n$&\multicolumn{1}{c}{$\sigma_0$}&$\sigma_2^{\rm
T}$&\multicolumn{1}{c}{$\sigma_2^{\rm D}$}&$\sigma_3^{\rm
T}$&\multicolumn{1}{c}{$\sigma_3^{\rm D}$}&$\sigma_3^{\rm
C}$&\multicolumn{1}{c}{$\sigma_{\rm tot}$}&\\\hline
  f&0&   2.8330 &     4.196$\times 10^{-3}$&   $7.327\times 10^{-4}$&   $-2.89\times 10^{-4}$&$-2.25\times 10^{-5}$& $-4.70\times 10^{-5}$&   2.8376 &\\
  ${\rm p}_1$&1&   3.7712 &   3.696$\times 10^{-3}$&   $1.364\times 10^{-3}$&   $-1.85\times 10^{-4}$&$-4.13\times 10^{-5}$& $-5.92\times 10^{-5}$&   3.7760 &\\
  ${\rm p}_2$&2&   4.9925 &   2.947$\times 10^{-3}$&   $1.750\times 10^{-3}$&   $-1.11\times 10^{-4}$&$-4.38\times 10^{-5}$& $-5.24\times 10^{-5}$&   4.9970 &\\
  ${\rm p}_3$& 3&   6.0592 &  2.470$\times 10^{-3}$&   $1.960\times 10^{-3}$&   $-7.72\times 10^{-5}$& $-4.16\times 10^{-5}$& $-5.34\times 10^{-5}$&   6.0635 &\\
  ${\rm p}_4$&  4&   7.0920 & 2.129$\times 10^{-3}$&   $1.953\times 10^{-3}$&   $-5.71\times 10^{-5}$&$-3.59\times 10^{-5}$& $-5.14\times 10^{-5}$&   7.0959 &\\
  ${\rm p}_5$& 5&   8.1662 &  1.856$\times 10^{-3}$&   $1.890\times 10^{-3}$&   $-4.33\times 10^{-5}$& $-3.04\times 10^{-5}$& $-4.44\times 10^{-5}$&   8.1698 &\\
  ${\rm p}_6$& 6&   9.2940 &  1.638$\times 10^{-3}$&   $1.849\times 10^{-3}$&   $-3.36\times 10^{-5}$& $-2.63\times 10^{-5}$& $-3.89\times 10^{-5}$&   9.2974 &\\
  ${\rm p}_7$& 7&   10.458 &  1.460$\times 10^{-3}$&   $1.697\times 10^{-3}$&   $-2.67\times 10^{-5}$& $-2.15\times 10^{-5}$& $-3.40\times 10^{-5}$&   10.461 &\\
  ${\rm p}_8$& 8&   11.641 &  1.317$\times 10^{-3}$&   $1.138\times 10^{-3}$&   $-2.16\times 10^{-5}$&$-1.30\times 10^{-5}$& $-3.12\times 10^{-5}$&   11.643 &\\
  ${\rm p}_9$& 9&   12.820 &  1.200$\times 10^{-3}$&  $-4.768\times 10^{-4}$&  $-1.80\times 10^{-5}$&  $4.99\times 10^{-6}$& $-3.03\times 10^{-5}$&   12.821 &\\
  ${\rm p}_{10}$& 10&   13.973 &  1.107$\times 10^{-3}$& $-4.344\times 10^{-3}$&$-1.53\times 10^{-5}$&  $4.18\times 10^{-5}$& $-3.13\times 10^{-5}$&   13.970 &\\
  ${\rm p}_{11}$&11&   15.092 &  1.030$\times 10^{-3}$&  $-1.138\times 10^{-2}$&$-1.32\times 10^{-5}$&   $1.02\times 10^{-4}$& $-3.21\times 10^{-5}$&   15.082 &\\
  ${\rm p}_{12}$&12&   16.197 &  9.630$\times 10^{-4}$&  $-2.066\times 10^{-2}$&$-1.15\times 10^{-5}$&   $1.72\times 10^{-4}$& $-3.04\times 10^{-5}$&   16.177 &\\
  ${\rm p}_{13}$&13&   17.312 &  9.018$\times 10^{-4}$&  $-3.214\times 10^{-2}$&$-1.01\times 10^{-5}$&  $2.51\times 10^{-4}$& $-2.69\times 10^{-5}$&   17.281 &\\
  ${\rm p}_{14}$& 14&   18.433 &  8.470$\times 10^{-4}$& $-5.165\times 10^{-2}$& $-8.90\times 10^{-6}$&  $3.80\times 10^{-4}$& $-2.34\times 10^{-5}$&   18.383 &\\
\hline
\end{tabular}\\
\end{center}
\label{Table-Sigma0-T23-D23-C-total-l2m1-ZAMS}
\end{table}
\clearpage
\begin{table}
\caption[]{Same as Table
\ref{Table-Sigma0-T23-D23-C-total-l1m1-ZAMS}, for f and p modes with
$(l,m)=(2,2)$ and $n=(0, \ldots ,14)$.}
\begin{center}
\begin{tabular}{rrrcrcrcrccccc} \hline
\multicolumn{1}{c}{$Mode$}&$n$&\multicolumn{1}{c}{$\sigma_0$}&$\sigma_2^{\rm
T}$&\multicolumn{1}{c}{$\sigma_2^{\rm D}$}&$\sigma_3^{\rm
T}$&\multicolumn{1}{c}{$\sigma_3^{\rm D}$}&$\sigma_3^{\rm
C}$&\multicolumn{1}{c}{$\sigma_{\rm tot}$}&\\\hline
 f&0&   2.7478 &     6.772$\times 10^{-4}$&   $-1.418\times 10^{-3}$& $-5.38\times 10^{-5}$&   $8.90\times 10^{-5}$&   1.07$\times 10^{-4}$&   2.7472 &\\
 ${\rm p}_1$& 1&   3.6591 &  1.204$\times 10^{-3}$&    $-2.159\times 10^{-3}$& $-8.78\times 10^{-5}$&  $1.30\times 10^{-4}$&   1.73$\times 10^{-4}$&   3.6584 &\\
 ${\rm p}_2$&  2&   4.8688 & 1.144$\times 10^{-3}$&   $-2.276\times 10^{-3}$& $-6.83\times 10^{-5}$&   $1.14\times 10^{-4}$&   1.96$\times 10^{-4}$&   4.8679 &\\
 ${\rm p}_3$&  3&   5.9318 & 9.859$\times 10^{-4}$&   $-1.848\times 10^{-3}$& $-4.97\times 10^{-5}$&   $7.88\times 10^{-5}$&   2.30$\times 10^{-4}$&   5.9312 &\\
 ${\rm p}_4$& 4&   6.9627 &  8.531$\times 10^{-4}$&    $-1.224\times 10^{-3}$& $-3.71\times 10^{-5}$&  $4.51\times 10^{-5}$&   2.36$\times 10^{-4}$&   6.9625 &\\
 ${\rm p}_5$& 5&   8.0358 &  7.468$\times 10^{-4}$&    $-9.157\times 10^{-4}$& $-2.83\times 10^{-5}$&  $2.95\times 10^{-5}$&   2.10$\times 10^{-4}$&   8.0359 &\\
 ${\rm p}_6$& 6&   9.1626 &  6.624$\times 10^{-4}$&    $-6.114\times 10^{-4}$& $-2.22\times 10^{-5}$&   $1.74\times 10^{-5}$&   1.91$\times 10^{-4}$&   9.1629 &\\
 ${\rm p}_7$& 7&   10.326 &  5.932$\times 10^{-4}$&    $-1.074\times 10^{-4}$& $-1.78\times 10^{-5}$&  $2.74\times 10^{-6}$&   1.73$\times 10^{-4}$&   10.326 &\\
 ${\rm p}_8$&  8&   11.508 & 5.350$\times 10^{-4}$&     $1.394\times 10^{-3}$& $-1.44\times 10^{-5}$& $-3.21\times 10^{-5}$&   1.66$\times 10^{-4}$&   11.510 &\\
 ${\rm p}_9$& 9&   12.687 &  4.848$\times 10^{-4}$&    $5.227\times 10^{-3}$& $-1.19\times 10^{-5}$&  $-1.10\times 10^{-4}$&   1.68$\times 10^{-4}$&   12.693 &\\
 ${\rm p}_{10}$&10&   13.840 &   4.406$\times 10^{-4}$&     $1.378\times 10^{-2}$& $-9.97\times 10^{-6}$& $-2.66\times 10^{-4}$&   1.81$\times 10^{-4}$&   13.854 &\\
 ${\rm p}_{11}$& 11&   14.959 &  4.036$\times 10^{-4}$&    $2.844\times 10^{-2}$&  $-8.48\times 10^{-6}$& $-5.10\times 10^{-4}$&   1.91$\times 10^{-4}$&   14.987 &\\
 ${\rm p}_{12}$& 12&   16.064 &  3.759$\times 10^{-4}$&     $4.673\times 10^{-2}$& $-7.37\times 10^{-6}$& $-7.82\times 10^{-4}$&   1.84$\times 10^{-4}$&   16.111 &\\
 ${\rm p}_{13}$& 13&   17.181 &  3.548$\times 10^{-4}$&  $6.849\times 10^{-2}$& $-6.51\times 10^{-6}$&    $-1.07\times 10^{-3}$&   1.65$\times 10^{-4}$&   17.249 &\\
 ${\rm p}_{14}$& 14&   18.305 &  3.364$\times 10^{-4}$&     $1.043\times 10^{-1}$& $-5.81\times 10^{-6}$& $-1.54\times 10^{-3}$&   1.47$\times 10^{-4}$&   18.408 &\\
\hline
\end{tabular}\\
\end{center}
\label{Table-Sigma0-T23-D23-C-total-l2m2-ZAMS}
\end{table}
\clearpage
\clearpage
\clearpage
\clearpage
\clearpage
\begin{table}
\caption[]{Values of zero-order eigenfrequency $\sigma_0$, total
frequency $\sigma_{\pm}$ corrected up to third order, total
third-order frequency correction
$\Delta\sigma_{\pm}=(\sigma_{\pm}-\sigma_0)$ due to rotation and
coupling, expansion coefficients $\mathcal{A}_1^{(\pm)}$ and
$\mathcal{A}_2^{(\pm)}$ normalized to 1, i.e.,
$\mathcal{A}_1^2+\mathcal{A}_2^2=1$, for selected pairs of
near-degenerate poloidal modes in a zero-age main-sequence star with
$M=12M_{\odot}$. All frequencies are in units of
$\sqrt{GM/R^3}=2.42\times 10^{-4}\, {\rm s}^{-1}$.}
\begin{center}
\begin{tabular}{ccccrrcrrcccccc} \hline
$m$&$l$&$n$&coupling&\multicolumn{1}{c}{$\sigma_0$}&\multicolumn{1}{c}{$\sigma_{\pm}$}&$\Delta\sigma_{\pm}$&\multicolumn{1}{c}{$\mathcal{A}_1^{(\pm)}$}&\multicolumn{1}{c}{$\mathcal{A}_2^{(\pm)}$}&\\\hline\hline
0&0&1&${\rm p}_1$&2.9941 &3.0027 &8.5940$\times 10^{-3}$&$0.99950$&$-0.03152$&\\
0&2&0&f&2.9199 &2.9262 &6.2499$\times 10^{-3}$&$0.04412$&$0.99903$&\\
\hline
0&0&3&${\rm p}_3$&5.1853 &5.1907 &5.3200$\times 10^{-3}$&$0.99719$&$0.07496$&\\
0&2&2&${\rm p}_2$&5.1189 &5.1266 &7.6955$\times 10^{-3}$&$0.08727$&$-0.99618$&\\
\hline
0&0&9&${\rm p}_9$&11.932 &11.934 &2.5947$\times 10^{-3}$&$0.99858$&$0.05319$&\\
0&2&8&${\rm p}_8$&11.775 &11.782 &7.0084$\times 10^{-3}$&$0.05738$&$-0.99835$&\\
\hline
\end{tabular}\\
\end{center}
\label{TwoModesCoupling-ZAMS}
\end{table}
\clearpage
\section{Concluding remarks}
The third-order effect of rotation on p and {\bf f} modes for a
uniformly rotating zero-age main-sequence star of mass 12~$M_\odot$
has been investigated. The third-order perturbation formalism
presented by Soufi et al. (1998) was used and revised because of
some misprints and missing terms in some of their equations.
Following Soufi et al. (1998), the Coriolis and
spherically-symmetric distortion effects were included in the
zero-order eigensystem. This yields eigenfrequencies $\omega_0$ of
eigenmodes which are no longer $m$-degenerate, even at zero order.
Furthermore this procedure enables one to obtain eigenfrequencies
with the required $\epsilon^3$ accuracy without the computation of
eigenfunction corrections at successive orders of $\epsilon$. The
zero-order eigenvalue problem was solved by pulsation code modified
in this manner. Numerical calculations of oscillation frequencies
were carried out for our selected model and second- and third-order
frequency corrections due to Coriolis, non-spherically-symmetric
distortion and Coriolis-distortion coupling were computed. For the
case of $m=0$ there is no first and third-order frequency
corrections. For the case of radial oscillation ($l=m=0$) the
second-order non-spherically-symmetric distortion is also zero and
only the second-order Coriolis contribution exist. We discuss the
validity of neglecting the surface terms which arise when the
density derivatives are removed through an integration by parts.
They become significant for higher-order modes, particularly in the
present model whose atmosphere is relatively thin. Coupling only
occurs between two poloidal modes with the same $m$ and with $l$
differing by 0 or 2.

We have carried out a careful comparison with the results of the
independent implementation of the third-order formalism by Soufi et
al. (1998). After taking into account the modifications discussed in
Appendix~\ref{Appendix-summary} the results of the two formalisms
for the combined second- and third-order corrections agree to within
a few per cent.

In a subsequent paper we intend to investigate numerically the
effect of rotation up to third order for a sequence of $\beta$
Cephei star models with uniform or radially varying rotation
profiles.

\begin{acknowledgements}
This work was supported by: the Department of Physics, University of
Kurdistan, Sanandaj, Iran; the Research Institute for Astronomy $\&$
Astrophysics of Maragha (RIAAM), Maragha, Iran; the Theore\-tical
Astro\-physics Center, a collaborative center between Copenhagen
University and Aarhus University, Denmark, funded by the Danish
National Research Foundation; and the Observatory of Paris-Meudon,
France; I wish to thank Prof. Christensen-Daslgaard for given the
model and his hospitality during the period that I visited Aarhus
University. I thank Profs. Goupil and Dziembowski for providing
valuable consultations. I also thank Dr. Reza Samadi for his
hospitality during the period that I visited the Observatory of
Meudon-Paris.
\end{acknowledgements}

\begin{appendix}

\clearpage
\section{Summary of modifications compared with the formulation of S98}
\label{Appendix-summary}
The third-order perturbation formalism for the influence of rotation
on stellar oscillations that is presented here is to a large extent
a re-derivation of the results of S98. We consider it necessary to
present some results of the derivations because of a number of
typographical errors and some missing terms in the paper of S98,
which might lead to confusion. For convenience we here summarize the
required corrections to that paper, and other modifications in the
present formulation:

\begin{itemize}

\item{} There are several corrections to S98, Eqs. (A10).
In the second of these equations the term $-h_{1}^{2}/\zeta$ is
mistakenly omitted (cf. Eq. (\ref{ytdrdifeq})). In the same equation
$U/(1-\sigma_r)$ must be used instead of $U$ in the radial case (cf.
Eq. (\ref{ytdrdifeqrad})). In the last of S98, Eqs. (A10), the term
in $v$ should be
$$
\left(\Lambda-\frac{UV_g}{1-\sigma_r}\right)v
$$
(cf. Eq. (\ref{wdrdifeq})).

\item{} In the definition of $\zeta$, S98 Eqs. (A11),
$C_r$ must be used instead of $C$ (cf. Eq.~(\ref{zetah1def})).

\item{}
In their Eq.~(A15), S98 neglected the factor $(g_{\rm e}/\tilde{g})$
as being $O(\Omega^2)$ and approximated $(g_{\rm e}/\tilde{g})v_k$
by $v_k$. We use the full expression in Eq.~(\ref{dfacdef}).

\item{} The factor $2J/\omega_0$ in Eq.~(A22) of S98 was
mistakenly omitted. The corrected form of the relation should be
\begin{equation}
\begin{array}{rl}
\langle\xi_{{\rm t}1q}\mid\tilde{\rho}\xi_{{\rm t}1k}\rangle=
\delta_{m_k,m_q}&\left(\displaystyle\frac{2J_{kq}}{\displaystyle\omega_0}\right)
\left(\frac{\displaystyle\bar{\Omega}}{\displaystyle\omega_0}\right)^2
[\delta_{l_k,l_q}K_1+ \\
&\delta_{l_k,l_q+2}K_{2kq}+\delta_{l_k,l_q-2}K_{2qk}^{\ast}] \; ,\\
\end{array}
\label{torcrotal}
\end{equation}
where $K_1$ and $K_{2kq}$ are given by Eq. (B5) in S98.
\end{itemize}

\subsection{Surface effects}\label{Appendix-surface}
We have considered two approaches for the computation of the
frequency corrections. In the first, the corrections are calculated
from Eqs. (B6)--(B7), (B19)--(B21), and (B25) of S98. In the second
which is used in present work, those coefficients are computed from
Eqs. (\ref{sigTsecondord}), (\ref{sigTthirdord}),
(\ref{sigDsecondord}), (\ref{sigDthirdord}), and
(\ref{sigCthirdord}).

\begin{figure}
 \includegraphics{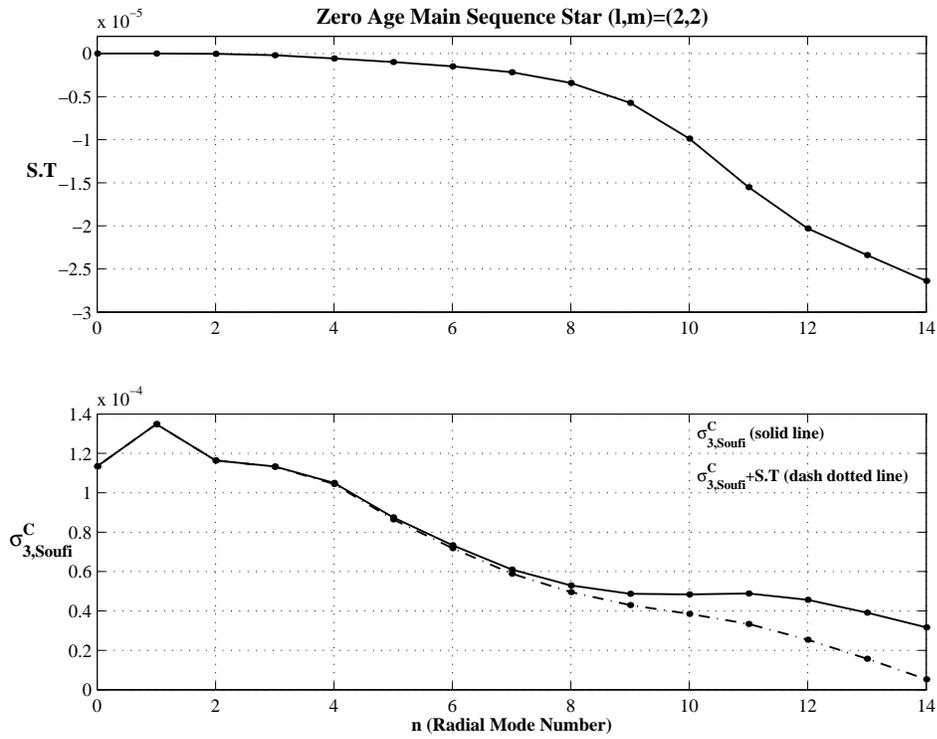}
      \vspace{16.3cm}
      \caption[]{Surface term and its effect on Eq. (B25) in Soufi et al. (1998)
      for p modes with $(l,m)=(2,2)$ and $n=(0, \ldots ,14)$
      for a zero-age main-sequence model.
       }
         \label{ZAMS-ST-STplusCc-l2m2}
   \end{figure}

There are two substantial differences between the equations used in
the two approaches. In the formulation of S98, the density
derivatives are eliminated through an integration by parts and the
resulting surface terms are ignored. For instance, in Eq. (B25) of
S98 the neglected surface term is
\begin{eqnarray}
{\rm S.T}= \frac{m}{J}\frac{\bar\Omega^3}{\omega_0} \left.
r^5u_2(1+\eta)(\delta_{l_kl_q}g_1+\mathcal{Q}_{kq2}g_2)
\tilde{\rho}\right|_{r=R} \; , \label{ST-B25}
\end{eqnarray}
where $g_1=C\sigma_0^2z_kz_q$ and
$g_2=C\sigma_0^2(y_kz_q+y_qz_k+3z_kz_q)$. In
Fig.~\ref{ZAMS-ST-STplusCc-l2m2}, the surface term given by Eq.
(\ref{ST-B25}) as well as its effect on the third-order distortion
and Coriolis coupling correction, $\sigma_{3,\rm Soufi}^{\rm C}$
(see Eq. (B25) in S98), are plotted against radial order. The plot
shows that for lower radial orders $n$ the effect of surface term is
negligible whereas for higher orders, it becomes important. Note
that for higher radial orders, the dominant term in surface term
comes mostly from $g_2$. Our study shows that for the model used in
the present paper the surface terms become considerable. {\bf
Because the boundary has been located at $X:=r/R=1$, where the
density is not exactly zero.}

\subsection{Using the approximation $z \simeq y_t/C\sigma_0^2$ in the third-order correction terms}\label{Appendix-approximation}

\begin{figure}
 \includegraphics{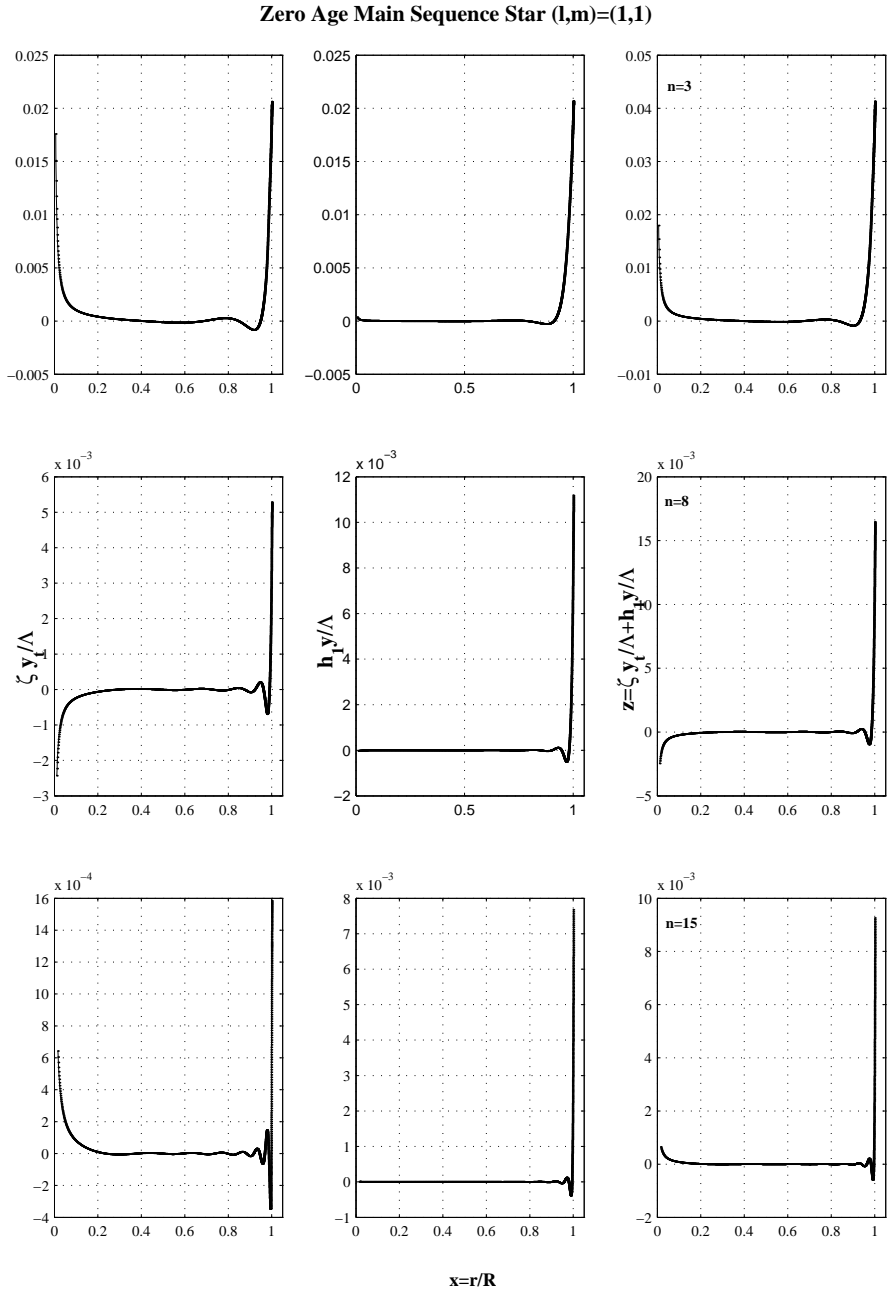}
      \vspace{19.7cm}
      \caption[]{The first $\zeta y_t/\Lambda$ (left),
the second $h_1y/\Lambda$ (middle), and the total terms (right) of
      zero-order horizontal component $z$ of eigenfunction against fractional radius $x=r/R$
      for selected p modes $(l,m)=(1,1)$ and $n=(3,8,15)$.}
         \label{ZAMS-FT-ST-Ztot-l1m1}
   \end{figure}

The other important difference which should be noted is that in the
S98 approach for computing the third-order correction terms the
approximation $z \simeq y_t/C\sigma_0^2$, which is valid for the
non-rotating case, is used. In the present paper, on the other hand,
the exact relation Eq. (\ref{ampleqn1}) is used. To investigate this
difference in detail, in Fig.~\ref{ZAMS-FT-ST-Ztot-l1m1} the two
terms included in Eq. (\ref{ampleqn1}) as well as the full relation
for $z$ are plotted for several p modes with $(l,m)=(1,1)$ and
$n=(3,8,15)$ in the 12$M_\odot$ ZAMS model considered here.
Figure~\ref{ZAMS-FT-ST-Ztot-l1m1} shows that from the centre to near
the surface the second term $h_1y/\Lambda$ is negligible compared
with the first term $\zeta y_t/\Lambda$. However, very close to the
surface the second term becomes important and with increasing radial
order $n$ it becomes dominant compared with the first term. Hence we
conclude that the approximation of neglecting the second term
everywhere, particularly near the surface, is not valid. We find
that the magnitude of this difference between the two approaches is
more significant than the magnitude of the difference due to of the
surface terms.

If we include the surface terms in the S98 approach and use the
approximation $z \simeq y_t/C\sigma_0^2$ in the present formulation
the results of the two numerical approaches are in good agreement.
For lower radial orders, the difference in the combined second- and
third-order frequency correction $\sigma_{\rm c}$ is then smaller
than one percent and increases slightly for higher orders.
\clearpage
\clearpage
\clearpage
\section{Different contributions of non-spherically-symmetric
distortion}\label{Appendix-Dkq}

The dimensionless off-diagonal terms $\bar{D}_{kq}$, Eq. (\ref{Dkq})
are split into five separate integrals for convenience:
\begin{eqnarray}
\bar{D}_{kq}^{(1)}=-\frac{\sigma_{\bar{\Omega}}^2}{2\bar{J}_{kq}}\int_0^1
dx \bar{\tilde{\rho}}x^4\Gamma_1\lambda_k\lambda_qu_2 \; ,
\end{eqnarray}
\begin{eqnarray}
\bar{D}_{kq}^{(2)}&=&\frac{\sigma_{\bar{\Omega}}^2}{2\bar{J}_{kq}}\int_0^1
dx \bar{\tilde{\rho}}x^4(1-\sigma_r)\{b_2-(A+V_g)u_2\}
\nonumber \\
&\times&\{y_qw_k+w_qy_k+(\bar{\Lambda}-3)(z_qv_k+v_qz_k)\} \; ,
\end{eqnarray}
\begin{equation}
\begin{array}{rl}
\bar{D}_{kq}^{(3)}&=
-\frac{\displaystyle\sigma_{\bar{\Omega}}^2}{\displaystyle
2\bar{J}_{kq}} \displaystyle\int_0^1dx \bar{\tilde{\rho}}x^4\Bigl\{
\frac{\displaystyle u_2}{\displaystyle 2} \times\Bigr. \\
&\times [z_q(\Lambda_q-\Lambda_k+6)(\lambda_k-(A+V_g)y_k) \\
&~~~~+(\lambda_q-(A+V_g)y_q)(\Lambda_k-\Lambda_q+6)z_k] \\
&+[b_2-(A+V_g)u_2+\frac{2}{3}(1+\eta_2)+ \frac{\displaystyle
1}{\displaystyle x}
\frac{\displaystyle d\bar{\phi}_{22}}{\displaystyle dx}] \\
&~~~~\Bigl.\times[\lambda_ky_q+y_k\lambda_q-2(A+V_g)y_ky_q]\Bigr\} \; , \\
\end{array}
\end{equation}
\begin{equation}
\begin{array}{rl}
\bar{D}_{kq}^{(4)}&=
\frac{\displaystyle\sigma_{\bar{\Omega}}^2}{\displaystyle
2\bar{J}_{kq}} \displaystyle\int_0^1 dx \bar{\tilde{\rho}}x^4
\left\{(1-\sigma_r)
\times \right. \\
&\times[b_2-(A+V_g)u_2][(U+\chi-4)y_ky_q \\
&~~~~~+(\lambda_q+\Lambda_qz_q)y_k+(\lambda_k+\Lambda_kz_k)y_q] \\
&-y_ky_q(A+V_g)[(2-\sigma_r)(b_2-(A+V_g)u_2) \\
&~~~~~~~~~+\frac{2}{3}(1+\eta_2)+\frac{\displaystyle
1}{\displaystyle x}
\frac{\displaystyle  d\bar{\phi}_{22}}{\displaystyle  dx}] \\
&-2b_2[y_ky_q+\frac{1}{4}y_kz_q(\Lambda_q-\Lambda_k+6) \\
&~~~~~~~~~\left.+\frac{1}{4}y_qz_k(\Lambda_k-\Lambda_q+6)]\right\} \; , \\
\end{array}
\end{equation}
\begin{equation}
\begin{array}{rl}
\bar{D}_{kq}^{(5)}&=
-\frac{\displaystyle\sigma_{\bar{\Omega}}^2}{\displaystyle
2\bar{J}_{kq}} \displaystyle\int_0^1  dx\bar{\tilde{\rho}}x^4
C(\frac{\displaystyle\sigma_{0k}^2+\sigma_{0q}^2}{\displaystyle 2})\times\\
&\times [b_2-(A+V_g)u_2][y_ky_q+(\bar{\Lambda}-3)z_kz_q] \; ,\\
\end{array}
\end{equation}
where $\bar{\Lambda}=(\Lambda_k+\Lambda_q)/2$ and
\begin{equation}
\begin{array}{rl}
\mathcal{Q}_{kk2}&=\frac{3}{2}(\beta_{k+1}^2+\beta_k^2)-\frac{1}{2}
=(l_k+1)\beta_k^2-l_k\beta_{k+1}^2\\
&=\frac{\displaystyle \Lambda_k-3m_k^2}{\displaystyle 4\Lambda_k-3}
\; .
\end{array}
\label{taukk2}
\end{equation}

\section{The hermiticity of an oscillating rotating system}
\label{Appendix-Hermiticity}
The hermiticity of the equations of oscillation in a rotating fluid
has already been investigated by Gough \& Thompson (1990) who
ensured the hermiticity for the rotating star by means of mapping
each point in the distorted model to a corresponding point in the
spherically symmetric stellar model. Lynden-Bell \& Ostriker (1967)
demonstrated that all of the linearized operators appearing in a
rotating system for a constant rotation profile are hermitian. Here
we extend the argument of Lynden-Bell \& Ostriker (1967) to the case
of a rotation profile that is a function of radius.

To study the hermiticity of the operator $\mathcal{L}$, Eq. (21) in
S98, we need to show only that the operator $A$, Eq. (22) in S98, is
hermitian. The other operators $B$, $D$ and $C$ which appear in Eq.
(22) of S98, also appear in the total coupling coefficient
$\mathcal{H}$ (see Eqs. (B1)--(B2) of S98), which is symmetric and
hermitian, as is shown in Sect. \ref{Hkq}.

Let $\xi$ and $\zeta$ be two displacement eigenvectors of an
oscillating rotating system. Then one can show that the operator
$A$, Eq. (22) of S98, satisfies the following relation
\begin{eqnarray}
\langle\zeta\mid A\xi\rangle=\langle A\zeta\mid\xi\rangle+{\rm S.T.}
\; , \label{Ahermstrt}
\end{eqnarray}
where the surface term (S.T.) is
\begin{equation}
\begin{array}{rl}
{\rm S.T.}=\oint_S&\left\{\tilde{p}'\zeta^\ast+
(\zeta^\ast\cdot\nabla\tilde{p})\xi+
(\Gamma_1\tilde{p}\nabla\cdot\zeta^\ast)\xi\right. \\
&+\tilde{\rho}(\tilde{\phi}'\zeta^\ast-\hat{\phi}'^\ast\xi) \\
&\left.+\frac{\displaystyle 1}{\displaystyle 4\pi G}
(\tilde{\phi}'\nabla\hat{\phi}'^\ast-\hat{\phi}'^\ast
\nabla\tilde{\phi}')\right\} \cdot d\mathbf{S} \; , \\
\end{array}
\label{allsurfterms}
\end{equation}
with $d\mathbf{S}=\mathbf{e}_rR^2\sin\theta d\theta d\phi$; here hat
and tilde indicate components of the eigenfunctions associated with
$\zeta$ and $\xi$, respectively. In particular, $\hat{\phi}'$ is
given by the Poisson equation, Eq.~(\ref{phi22term}), for
eigenvector $\zeta$. Following Unno et al. (1989), the relevant
boundary conditions for an oscillating rotating star are
\begin{equation}
\tilde{p}=\tilde{\rho}=0 \; ~~~~~~{\rm at}~r=R \; ,
\label{zerpressurf}
\end{equation}
\begin{equation}
\tilde{p}'=0 \; ~~~~~~{\rm at}~r=R \; , \label{zerperpressurf}
\end{equation}
\begin{equation}
\frac{d\phi'}{dr}+\frac{(l+1)}{r}\phi'=0 \; ~~~~~~{\rm at}~r=R \; .
\end{equation}
The first and second boundary conditions, Eqs.~(\ref{zerpressurf})
and (\ref{zerperpressurf}), result in removing all terms in
Eq.~(\ref{allsurfterms}) that include $\tilde{p}$ and $\tilde{p}'$,
and also all terms that include $\nabla\tilde{p}$ because they are
proportional to $\tilde{\rho}$ (see Eqs.~(12) and (19) in S98). The
last boundary condition eliminates the last surface term in
Eq.~(\ref{allsurfterms}). Hence all the surface terms go to zero at
the surface of the star and therefore ${\rm S.T.}=0$, with the
result that Eq.~(\ref{Ahermstrt}) becomes
\begin{eqnarray}
\langle\zeta\mid A\xi\rangle=\langle A\zeta\mid\xi\rangle \; ,
\end{eqnarray}
and therefore the operator $A$ is hermitian.

We note that from a theoretical point of view, the density vanishes
exactly at the surface, whereas for a realistic model obtained from
numerical calculations the density is not exactly zero at the
surface. Hence from a numerical point of view the surface terms are
not exactly zero; consequently, whether the surface term can be
removed or not depends on the required accuracy. For instance, we
saw before that the effect of surface terms is not negligible when
the cubic frequency corrections are computed.
\end{appendix}


\label{lastpage}

\end{document}